%% file: ulg101.tex
\documentclass[12pt,a4paper]{article}
\usepackage[utf8]{inputenc}
\usepackage[T1]{fontenc}
\usepackage{amsmath,amssymb,amsfonts,mathtools}
\usepackage{bm}               
\usepackage{slashed}          
\usepackage{graphicx}
\usepackage{booktabs}         
\usepackage{multirow}
\usepackage{array}
\usepackage{dcolumn}          
\usepackage{xcolor}
\usepackage{comment}
\usepackage{tikz}             
\usepackage{siunitx}          
\usepackage{hyperref}
\usepackage{cleveref}         
    \let\MeV\relax  \let\GeV\relax
 \let\fm\relax

\providecommand{\MeV}{\,\mathrm{MeV}}
\providecommand{\GeV}{\,\mathrm{GeV}}

\providecommand{\fm}{\,\mathrm{fm}}

\providecommand{\alphas}{\alpha_s}

\providecommand{\dd}{\mathrm{d}}                           

\newcommand{\beq}{\begin{equation}}
\newcommand{\eeq}{\end{equation}}
\newcommand{\bea}{\begin{eqnarray}}
\newcommand{\eea}{\end{eqnarray}}
\hypersetup{
    colorlinks=true,
    linkcolor=blue,
    citecolor=red,
    urlcolor=teal 
}
\title{\boldmath Ultralight Glueball}
\author{Shi-Yuan Li\\
\small Institute of Theoretical Physics, Shandong University, Jinan 250100, PR China\\
\small lishy@sdu.edu.cn}
\date{\today}

\begin{document}

\maketitle

\begin{abstract}
Recent experimental progress and lattice studies have primarily focused on the heavy glueball sector.
Yet the existence or non-existence of an ultralight glueball with mass
below $2m_\pi$, possibly even below $m_\pi$ itself, carries implications on confinement far beyond spectroscopy.
Contrary to the heavy candidates  inevitably  mixing with flavour-singlet quark components,
such an almost pure-glue bound state   would reveal a soft  confinement scale other than  $m_\pi$,
offering a new probe of the infrared structure of QCD and the origin of its mass gap.
An ultralight glueball can only   weakly couples   with the nucleon    via spatial overlap of gluon cloud,
evading standard hadron searches. Its internal gluonic parton structure could be resolved only in sufficiently energetic collisions.
We investigate its imprints in phenomenology of various  arenas, such as nuclear force and   structure, elastic scattering and  multi-production (especially  cosmic-ray extensive air showers) at high-energy, inclusive   $J/\psi$   radiative decay,
as well as the r\^ole in early universe and dense quark matter.
We note that current experiments have not excluded such a glue-hadron. Only if forthcoming experiments continue to constrain its coupling  downward  effectively to zero, one may tentatively conclude that it is excluded.
On the positive side, two possible signals can reveal its presence:  First, a near-ground excess of $\pi^0$-like signals   in extensive air shower arrays could indicate its production in cosmic-ray interactions. Second,  its distinctive spectral feature should appear near the upper
endpoint of the photon energy spectrum in radiative \(J/\psi\) and \(\Upsilon\) decays,  potentially accompanied by
a coincidence signal in a backward calorimeter,  a characteristic signature of its gluonic and hadronic properties.
We also emphasize that even  exclusion  for the  ultralight glueball would not render the energy region below $2m_\pi$ hadronic desert.
It would instead motivate an alternative  possibility in which gluonic degrees of freedom persist through a continuous,
quark-coupled infrared structure, potentially admitting an approximately scale-invariant, unparticle-like description.

 In one word, the region below the two-pion threshold, extending down to zero energy, is far from being a hadronic desert;
rather, it is an important window into the low-energy gluonic structure of QCD and the mechanism of color confinement.
 The ultralight glueball is a natural probe of this
energy window: its discovery would reveal a new
low-energy gluonic excitation in full QCD, while
its exclusion would establish quantitative constraints
on how strongly a sub-threshold gluonic state can
couple to ordinary hadronic matter, thereby providing
an important benchmark for models of the low-energy
gluonic sector.
\end{abstract}

\input{Introduction}         
\input{partA}
\input{partB}       
\input{partC}
\input{partD}          %

\bibliographystyle{unsrt}
\bibliography{glbrefs}

\end{document}

%% file: Introduction.tex

\section{Introduction}
\label{sec:intro}

Glueball is the colour singlet bound state of gluons. It is  a distinctive prediction of QCD as  a non-Abelian Yang-Mills gauge theory in which gluons carry colour charges and interact strongly with each other.   The experimental search focused on the heavy sector and obtained  evidence recently, while lattice QCD and QCD sum rules have long predicted masses well above $1\GeV$.  However, ``heavy'' does not mean ``only heavy''.  Gluons are massless in the QCD Lagrangian and no constituent-gluon model exists to rival the constituent-quark picture.
Hence comes the question what  the r\^ole of the gluon field in low energies  smaller than 1GeV is, or whether it is still an independent degree of freedom.
To answer these questions, light and ultralight glueball should be studied, even just to establish reliable exclusion limits. The ultralight glueball with mass  below $2m_\pi$, even below $m_\pi$ itself \footnote{To clarify the terminology, we refer the glueball with mass in the range heavier than  $2m_\pi$ but lighter than the bound states predicted by lattice QCD as the light glueball. This  trivial kind shares similarity with the heavy ones, and also mixes with quark components. On the other hand,  the glueball with mass smaller  than $2m_\pi$ are referred as  ultralight. In the following we only concentrate on such ultralight ones.
As a very special case,
an ultralight glueball with a mass almost degenerate
with $m_\pi$ can mix with the $\pi^0$ through isospin-breaking effects
parametrically suppressed by $(m_d-m_u)/\Lambda_{\rm QCD}$.
This possibility is generally neglected in the present work, except where
strictly necessary to address, as it does not alter the kinematic isolation
or the weak coupling that underpin our analysis. Another special case is that the ultralight glueball mass lies between $2m_\mu$ and $2 m_\pi$. It can decay into muon pair, which is also neglected in this paper. So in the following discussion one can generally  consider the ultralight glueball mass as smaller than $2m_\mu$.}, is the only clean `glue' bound state, since it does not mix with flavour-singlet meson of quark components with similar quantum number, while this mixing prevents heavy glueballs from being isolated as pure states. Most importantly, such ultralight glueball is colour 'insulator',  \emph{weakly} interacts with hadrons (only by gluon field overlap). This is contrary to all the other handrons (only  in hard  scatterings its partonic structure  manifests as general hadrons).  Hence   it can  evade  standard hadron searches, and is  missed  by the present experiments.
Such an ultralight and almost pure-glue state would reveal a soft confinement scale other than the pion mass, and would  inform us about the long-wavelength structure of the QCD vacuum.  It can offer far deeper insight into QCD confinement than a mere spectroscopic addition. Even if it does not exist, the experimental constraints we derive constitute a direct probe of the gluon field in this infrared regime.
From these motivations,
 in the following sections
of this paper,  we study the imprints of the ultralight glueball  in phenomenology of various arenas and conclude that it  is not yet excluded. We also suggest  further necessary theoretical and experimental studies.  In the remainder of this section,  we   explain  the above motivations in details.


\subsection{The heavy glueballs have  experimental and theoretical evidences} 
\label{subsec:heavy_anchored}

  Using the world's largest $J/\psi$ sample of $\sim 10^{10}$ events at BEPCII, the BESIII Collaboration have determined the spin-parity of the $X(2370)$ resonance to be $J^{PC}=0^{-+}$ with mass
\beq
M_{X(2370)} = 2395 \pm 11\,(\text{stat})^{+26}_{-94}\,(\text{syst})\MeV/c^2 ,
\eeq
in  agreement with lattice-QCD expectations for the lightest pseudoscalar glueball~\cite{Morningstar:1999rf,Athenodorou:2020kdu,Vadacchino:2023vnc}.  Subsequent BESIII work established the flavor-singlet nature of $X(2370)$ through the absence of $K^*\bar K$ decays (forbidden for a $0^{-+}$ flavor singlet by generalized $G$-parity) and observed new decay modes $K_S^0K_S^0\pi^0$, $\pi^0\pi^0\eta$, and $a_0(980)^0\pi^0$ etc..  At ICHEP 2026, BESIII presented the complete evidence chain, mass, $J^{PC}$, flavor singlet, decay pattern, making $X(2370)$ the clearest glueball candidate to date~\cite{BESIII:2026mvn,BESIII:2026rzt,Egorov:2026yzz}.

For the scalar channel, lattice QCD consistently places the lightest $0^{++}$ glueball at
\beq
m_{0^{++}} \simeq 1.5\text{--}1.8\GeV ,
\eeq
with quenched and unquenched simulations in tight agreement~\cite{Morningstar:1999rf,Gregory:2012hu,Chen:2022asf}.  The pseudoscalar $0^{-+}$ ground state lies higher, $m_{0^{-+}}\simeq 2.3\text{--}2.6\GeV$.  A recent lattice study of the $0^{++}\to\phi\gamma$ radiative width further tightens the identification with $f_0(1500)/f_0(1710)$ mixing~\cite{Gui:2024glueball}.

QCD sum rules have been used to expalain the BES discovery (e.g., \cite{Wang:2026iyq}; using the gluonic currents
\bea
j_s(x) &=& \alphas G^a_{\mu\nu}(x) G^{a\mu\nu}(x) \quad (0^{++}), \\
j_{ps}(x) &=& \alphas G^a_{\mu\nu}(x) \tilde G^{a\mu\nu}(x) \quad (0^{-+}),
\eea
span a wider range.  Early works (Novikov et al., Narison) gave $0.7\text{--}0.9\GeV$ for the scalar; including radiative corrections and fermion loops drives the central value to $1.25\text{--}1.7\GeV$ ~\cite{Novikov:1980uj,Narison:1983di,Bagan:1994sg,Huang:1998wj,Yuan:2009zz}.  With instanton effects included, a scalar possibly  has light mass  of $\sim 400\text{--}600\MeV$,  but dropping gluonic condensates in favor of instantons removes this light pole, shifting solutions above $1.4\GeV$~\cite{Kisslinger:2001pk,HarnettSteele:2001}.  No standard sum-rule analysis produces a \emph{stable isolated pole below $\sim 500\MeV$}.

Thus the lightest glueball states predicted by standard lattice and sum-rule methods are heavy. At the same time, it is very natural for   a flavor-singlet hadron in the $2\text{--}3\GeV$ window  has a large part of gluon component provided that it is excluded from a multiple quark state \cite{Ben:2026afy,Peng:2026yol}. If QCD vacuum  supports such a prominent gluonic component in a heavy hadron, long wave gluonic fluctuations may also exist. The question addressed in this paper is whether such fluctuations can organize themselves into a low mass colour-singlet bound state.  

\subsection{``Heavy'' need not mean ``only heavy''}
\label{subsec:not_only_heavy}


Taken together, the  properties  of $X(2370)$ measured by BESIII, the stability of
the $0^{++}$ lattice spectrum across quenched and dynamical simulations,
and the persistence of a $\sim1.5\GeV$ glueball signal in QCD sum rules
constitute 
the evidence of the  heavy scalar/pseudoscalar glueball.
But one can not extrapolate further and assert that all glueballs
must live in the $1.5\!-\!3 \GeV$ window.
Lattice variational studies converge rapidly for local interpolating fields
e.g., $\mathcal{O}_i\sim G^2,\,(DG)^2$, but this convergence guarantees a
variational lower bound only for states with non-negligible overlap onto
that basis.
An ultralight glueball, overwhelmingly composed of long-wavelength zero modes 
of the gluon field, has a Compton wavelength $\lambda_G\sim 1/m_G$ that
far exceeds the localisation scale of these operators.
Its overlap with such local interpolators is therefore correspondingly
suppressed, and it need not appear as a clean plateau.
The spatial extent of existing lattice volumes is typically a few femtometres, whereas the Compton wavelength of a sub-pion state can be larger. Whether such a state can be accommodated within current lattice boxes, or would simply be truncated by the finite volume, remains an open question.

An interesting possibility arises in lattice QCD at sufficiently heavy pion masses. In a two-flavor lattice QCD study, \cite{Sun:2018fha}  extracted the scalar gluonic state using operators constructed from Wilson loops. At $(m_\pi \simeq 938)$ MeV, they obtained a mass of $(M_{0^{++}}=1417(30))$  MeV, substantially below the two-pion threshold $(2m_\pi \simeq 1876)$ MeV. Although glueball-meson mixing and quark-mass dependence remain to be clarified, this result can not be straightforwardly extrapolated to physical pion mass. But it  provides an interesting motivation for investigating the possible existence and properties of sub-threshold gluonic states in the confining vacuum.

QCD sum rules stabilise at $s_0\sim 3\,\GeV^2$ because the spectral integrals
are weighted by powers of $s$; a narrow sub-$\pi$ pole contributes a
negligible weight and can easily be missed once the continuum model
saturates the moments.
%


In the quark sector we possess the constituent-quark picture and chiral
symmetry as guiding rails; for gluons there is no analogue of a controlled
non-relativistic potential model grounded in first principles.
Gluons are massless in the Lagrangian; any effective constituent-gluon mass,
glueball scale, or confinement gap is an entirely non-perturbative output of
the QCD vacuum, about which we have no analytic control.
That the main glueball trajectory sits near $2\GeV$ does not logically prevent
the vacuum from supporting another, far softer collective mode of the gauge
field with size $\lambda\sim1/m_G >\Lambda_{\rm QCD}^{-1}$ puts it
outside the reach of local interpolating fields %
\footnote{%
The dilaton is the (pseudo-)Nambu--Goldstone boson associated with the
breaking of scale symmetry: it is strictly massless when scale symmetry is
spontaneously broken, and acquires a mass only through explicit breaking
($\beta(g)\neq0$, $m_q\neq0$).  Its mass and coupling are governed by
$\langle\theta^\mu_\mu\rangle$, not by the gluon-field binding mechanism.
Consequently, a dilaton need not be a glueball, and dilaton-search null
results are logically independent of the question addressed here---whether a
kinematically isolated gluon bound state can exist below $2m_\pi$.%
}.

Experimentally, BESIII triggers on prompt photons; a glueball with
$m_G< 2 m_\pi$ has its strongest accessible decay channels either
kinematically closed or severely phase-space suppressed, while leptonic
modes are helicity and mass suppressed.
This places the ultralight glueball in a kinematic desert where standard
analyses simply do not look. This  consequence of kinematic quarantine will be further discussed
the following
subsections on weakly interacting and section on $J/\Psi$ and $\Upsilon$ decay.



\subsection{Kinematic quarantine: the ultralight regime as the only
clean laboratory for the pure-glue hypothesis}
\label{subsec:clean_lab}
A physical hadron is an eigenstate of the QCD Hamiltonian and need not
be a pure Fock component. When a glueball basis state $|G\rangle$ and a flavour-singlet
$\bar q q$ meson state $|M\rangle$ share the same quantum numbers as $J^{PC}$, isospin,
$G$-parity, etc., and are quasi-degenerate in mass, the Hamiltonian is non-diagonal in
this subspace and the physical eigenstates are mixed,
\beq
|\psi\rangle = \cos\theta\,|G\rangle + \sin\theta\,|M\rangle .
\label{eq:mixed_state}
\eeq
For the heavy scalar glueballs in the $1.4$--$1.8\GeV$ region this is
precisely the case: the $0^{++}$ glueball is degenerate with members of
the scalar $\bar q q$ nonet, and the observed $f_0(1370)$, $f_0(1500)$
and $f_0(1710)$ are mixed eigenstates \cite{AmslerClose:1995,Close:2002scalar,Giacosa:2005qr}.
Crucially, a heavy glueball can therefore never be proven pure.  Any state
visible in detectors is already an eigenstate of the full Hamiltonian, and
its decay pattern reflects the superposition \eqref{eq:mixed_state}, not
the pure glue component.
The BESIII identification of $X(2370)$ as a glueball-dominated $0^{-+}$
state  is inferred to be $\sim80$--$90\%$ gluonic with a
few-degree $\bar q q$  admixture
\cite{Morningstar:1999rf,BESIII:2026mvn}.

Below $2m_\pi$ the situation changes qualitatively.
The physical eigenstate of the full Hamiltonian can still be written as
a superposition of the same form \eqref{eq:mixed_state}, but now the
relevant $\bar q q$ components can only be \emph{off-shell}: there is
no on-shell scalar $\bar q q$ state with mass below $2m_\pi$ into which
the glueball can mix, with the pion  the lightest $\bar q q$ hadron
by a wide margin.  This is the very reason for defining the ultralight
regime. 
%
With the strong-decay channels kinematically closed, the glueball cannot
access the quark-hadron continuum through ordinary decay.   
%
A strong virtual $q\bar q$ loop would also have observable consequences.
Because the quark and antiquark carry electric charge, such a loop would
mediate radiative transitions, in close analogy to $\pi^0\to\gamma\gamma$,
whose large width and readily detected final state make it one of the
best-measured particle signal.
The absence of any comparable electromagnetic signal below $2m_\pi$
therefore constrains strong virtual $q\bar q$ coupling in this regime (though it does not rigorously exclude the
glueball).  In this sense, the weakness of the glueball--quark coupling below
$2m_\pi$ is not assumed but inferred from the absence of an electromagnetic
signal that a strong coupling would inevitably produce.

The ultralight glueball is therefore kinematically quarantined from the
quark sector: its purity is not an assumption but a consequence of
spectral isolation, \emph{because there is no hadron in this region to
superimpose with}.
Only by confirming or excluding an ultralight glueball can one settle
whether a pure-glue bound state is a viable object in QCD.
The heavy sector cannot answer this question, because data cannot
distinguish a pure heavy glueball from a mixed one; the ultralight window
is the only regime where purity is enforced by kinematics rather than
assumed.  It is therefore the only clean laboratory for the pure-glue
hypothesis.

This kinematic quarantine has a direct consequence for detectability:
the same spectral isolation that protects the purity of the ultralight
state also suppresses its interactions with ordinary hadrons, making it a
weakly coupled QCD object.

\subsection{Colour insulation and the natural weakness of pure-glue
coupling}
\label{subsec:colour_insulation}

The weakness of the coupling of an ultralight glueball is the most
important ingredient that makes the whole enterprise consistent, and is
the key pillar of the study in this paper.  We place this subsection here
because it follows naturally from the kinematic and symmetry structure of
QCD in the ultralight window addressed above.

As argued in \S\ref{subsec:clean_lab}, a glueball below $2m_\pi$ has no
on-shell $\bar q q$ component to mix with, so the quark-exchange diagrams
that generate hadronic-strength couplings simply do not exist.  The
dominant strong-interaction mechanism that produces order-unity couplings
is therefore structurally absent.
For a heavy glueball such as $X(2370)$, even a few-percent $\bar q q$
admixture opens quark-exchange channels analogous to those mediating the
nuclear force; for an ultralight glueball this quark exchange channel is closed by
kinematics, not by a symmetry.

 The only residual interaction persists is
through the spatial overlap of the gluon field of the ultralight glueball with that
(gluon cloud) of the other hadron.  This is not a consequence of mixing,
but of the fact that both objects contain coherent gluon distributions.
Because the glueball is a flavour singlet and a colour insulator, this
overlap does not involve any quark bilinear vertex; it is a direct,
flavour-blind interaction between two gluon-rich configurations.

Crucially, this overlap does not benefit from any of the mechanisms that
enhance ordinary hadronic forces.  There is no light Goldstone boson
(such as the pion) to mediate a long-range force, nor any nearby
resonance to provide a dynamical enhancement.  The interaction is
therefore short-ranged and unenhanced, relying solely on the local
coincidence of two colour-singlet gluon fields.


It is important to distinguish the '\emph{internal} strength' from the
'\emph{external} weakness'.  Two glueballs (including glue-rich hadron) can
have  a large  gluon field overlap and hence a strong coupling
in direct analogy to pion
nucleon 'overlap' of quark field, which leads to a strong nuclear force. %
What kinematic quarantine
suppresses is therefore  not the internal glue-rich overlap dynamics, but only the coupling of
the ultralight glueball  to the \emph{quark} sector:  No quark-rich component as investigated above,
so it remains weakly coupled to ordinary quark-hadrons (among which the nucleon is the main component of any detector).  In this sense
the ultralight glueball lives in a self-interacting but externally
insulated world, i.e.,  strongly coupled among gluon-rich objects, weakly
coupled to the quark-hadrons.

\paragraph{Here we would like to point that  Electromagnetic effects are even weaker.}
To confirm that the colour-driven estimate above is not undermined by
electromagnetism, we compare the glueball with a quark nugget, a
colour-neutral chunk of quark matter with a globally neutral but locally
charged surface layer \cite{Witten:1984rs}.
Such an object develops a Debye-screened electromagnetic profile and a
finite dipole polarisability, giving a residual nugget--nugget
electromagnetic force of order $10^{-3}$ of the nucleon--nucleon
Coulomb force.
An ultralight glueball is even more inert: it has no net charge
distribution, no surface layer, and no permanent multipole moments in its
ground state.  Any electromagnetic coupling must therefore proceed
through virtual fluctuations of the quark-gluon sea.
The glueball--nugget comparison therefore confirms that the weakness
identified in (b) is not an artefact of neglecting photons.

The glueball--nucleon coupling is thus weak for three structural reasons:
no quark exchange, because it is kinematically closed; the flavour-blind
direct gluon-field overlap is unenhanced; and electromagnetic effects are
negligible compared even with the weakly coupled quark nugget.
This weakness is not an assumption, nor a fine-tuning, but the expected
behaviour of a flavour-singlet, colour-insulating collective mode in the
ultralight window.
Quantitatively, this places the glueball--nucleon coupling three orders
of magnitude below $g_{\pi NN}$, as will be derived from the gluon-cloud
spatial overlap in the next subsection.


%

\subsection{Quantitative estimate of the glueball--nucleon coupling}
\label{subsec:coupling_estimate}
 %
%
%
  Now we establish a conservative upper-bound on how strong the (ultralight) glueball--nucleon coupling can
reasonably be,  taking  a scalar glueball as example at first.
The real   coupling  could  be far smaller even experimentally inaccessible.
As a matter of fact, the main purpose of this whole paper is to  study various constrains,
and the results show that such a weak coupling ultralight glueball can exist.

  The preceding discussion rules out quark exchange and electromagnetic
multipoles as possible origins of a glueball--nucleon interaction.
What remains is the spatial overlap of the nucleon's gluon cloud with the
glueball's gluon field.  At distances of the nucleon radius and larger
this overlap is a low-energy phenomenon and cannot be described by
perturbative gluon exchange.
  We therefore work in an effective field theory framework and employ the
Lorentz-scalar, flavour-singlet, colour-singlet operator
$\mathcal G\equiv\alpha_s G^a_{\mu\nu}G^{a\mu\nu}$.
In the chiral limit the trace anomaly reduces to
$\theta^\mu_\mu=(\beta/2g)\,G^a_{\mu\nu}G^{a\mu\nu}\propto \mathcal G$,
which serves as the lowest-dimensional operator available to mediate the
overlap of two gluon fields at low energy.
The factor $\alpha_s$ here is not a running coupling in the perturbative
sense; it is the renormalisation convention required to make the gluon
field strength enter physical matrix elements.  We introduce the effective
vertex
\beq
\mathcal L_{\rm eff} \supset C_G\,\frac{\mathcal G}{\Lambda^3}\,\bar N N ,
\label{eq:effvertex}
\eeq
where $\Lambda$ is the ultraviolet threshold of this low energy confined-phase
effective theory, and $C_G$ is a dimensionless $\mathcal O(1)$ parameter
that encodes the non-perturbative overlap dynamics (see below).

 The ultralight scalar glueball $|G\rangle$ is the state created by
$\mathcal G$.  In covariant normalisation we define
\beq
\langle 0|\mathcal G(0)|G(\vec p)\rangle = m_G^2 f_G ,
\label{eq:Ginterp}
\eeq
with $f_G$ the glueball decay constant,  analogous to that of pion
$f_\pi\simeq130\MeV$.  The corresponding operator replacement is
\beq
\mathcal G(x) \;\longrightarrow\; m_G^2 f_G\, G(x) ,
\label{eq:op_to_field}
\eeq
where $G(x)$ is the propagating scalar glueball field.  Substituting
Eq.~\eqref{eq:op_to_field} into Eq.~\eqref{eq:effvertex} yields the
Yukawa vertex
\beq
\mathcal L_{\rm eff} \supset g_{GNN}\, G\, \bar N N,
\qquad
g_{GNN} \equiv C_G\, \frac{m_G^2 f_G}{\Lambda^3} .
\label{eq:gGNNdef}
\eeq
%

The denominator $\Lambda^3$ supplies the three powers of mass required to
match the dimension-four operator $\mathcal G$ to the Yukawa term.
 $\Lambda$ marks the 'ultraviolet' scale, above which the
 'gluon particle' degrees of freedom appear rather than the single collective glueball.
   It should therefore be bounded from below by the
lightest hadron that is \emph{not} lightened by the pseudo-Goldstone
mechanism.  The pion and the kaon are pseudo-Goldstone bosons;
and excluding them,  the lightest
hadronic states lie in the $500\MeV$ range (e.g., $\rho(770)$, $a_0(980)$,
$f_0(980)$, $K^*(892)$).  So  $\Lambda\gtrsim500\MeV$ is a very
conservative bound.  We adopt $\Lambda=500\MeV$ below, which makes the
resulting $g_{GNN}$ a conservative upper limit: a larger $\Lambda$ would
only make the coupling smaller.

For the glueball decay constant we take $f_\pi\simeq130\MeV$ as a
conservative upper bound.  The decay constant sets the strength of the
coupling to hadronic channels and hence controls the width.  Because an
ultralight glueball below $2m_\pi$ has no open strong-decay channel and
must decay only through electromagnetic or higher-order processes, its
lifetime exceeds that of the $\pi^0$ (whose principal mode is
$\pi^0\to\gamma\gamma$).  Its decay constant is therefore bounded by
$f_G\lesssim f_\pi$.

For the dimensionless overlap-strength parameter $C_G$, as argued in
\S\ref{subsec:colour_insulation}, the ultralight glueball lacks the
resonant and long-range enhancement mechanisms that amplify conventional
hadronic couplings (such as the $g_{\pi NN}$ enhancement by the
$\Delta$ resonance).  Barring such an enhancement, there is no dynamical
reason for $C_G$ to exceed order unity.  We therefore take
$C_G\sim\mathcal O(1)$ as a conservative reference, with the
understanding that any $C_G\gg1$ would require an unspecified dynamical
enhancement that the kinematic structure of the ultralight window does not
provide.

With the illustrative representative mass $m_G=100\MeV$, $f_G\sim f_\pi\simeq
130\MeV$, $\Lambda=500\MeV$ and $C_G\sim\mathcal O(1)$,
\beq
g_{GNN}=C_G\,\frac{m_G^2 f_G}{\Lambda^3}
\;\sim\; 10^{-2} ,
\label{eq:gGNNfinal}
\eeq
to be compared with the pion--nucleon Yukawa coupling
$g_{\pi NN}\simeq13$.  The (ultralight) glueball--nucleon coupling is therefore about
three orders of magnitude smaller,
\beq
\frac{g_{GNN}}{g_{\pi NN}}\sim\frac{10^{-2}}{13}\sim10^{-3} .
\label{eq:gGNNratio}
\eeq
The suppression has two independent origins: the overlap factor
$m_G^2 f_G\sim(0.1\GeV)^2(0.13\GeV)\sim10^{-3}\GeV^3$, and the
denominator $\Lambda^3\sim(0.5\GeV)^3\sim10^{-1}\GeV^3$.
Because $g_{GNN}\propto m_G^2$, lighter glueballs are more weakly
coupled.   

 The same reasoning applies to the pseudoscalar channel.  There the
mediating operator is $\operatorname{tr}F\tilde F=
\epsilon_{\mu\nu\rho\sigma}G^{\mu\nu}G^{\rho\sigma}$.  The scalar
channel contracts two field strengths directly; the pseudoscalar does
so through the four-dimensional totally antisymmetric tensor.  Both are
contractions of two field strengths, so both operators have mass
dimension four, and the  suppression factor in the effective vertex is
therefore the same in the two channels. If the overlap parameter $C_P$ is of the same order as
$C_G$,  the above estimation is the same.
Their detailed quantitative
differences are the subject of the  following sections.  

Comparing with the heavy glueball candidates mentioned above
\cite{Morningstar:1999rf} which inevitably mixing and hence strongly interacting  with the
quark sector,   ultralight glueball in the window below $2m_\pi$ is
complementary.  It concerns a pure-gluon
excitation whose very existence would demonstrate that the gluon sector
retains an independent long-wavelength degree of freedom below the chiral
scale.
The operators $\operatorname{tr}F^2$ and $\operatorname{tr}F\tilde F$
are orthogonal to quark-bilinear meson operators in the chiral limit,
yet if the nucleon can contains a gluon cloud,  its overlap with the
glueball field supplies the portal.  The coefficient $C_{G,P}$ in
Eq.~\eqref{eq:gGNNdef} is therefore the non-perturbative overlap
strength, not an arbitrary coupling constant.  We take them
$\sim\mathcal O(1)$ as a conservative reference; barring a dynamical
enhancement that the ultralight window does not provide. 
Conversely,  
 if  the
overlap vanishes,  this can happen either because the gluon field has
been removed from the infrared by a low-energy cutoff, the possibility
envisaged in \S\ref{subsec:scale_anxiety}, or because no such glueball
exists at all.  A vanishing coupling therefore does not merely mean that
the interaction is weak.  It may indicate that the ultralight glueball
itself is absent.

The remainder sections of the  paper investigate phenomenologically
 the upper bound on $\alpha_G$
quantitatively (or finds a process that forces the coupling to vanish,
but no found).  The scalar and pseudoscalar channels are treated
separately, and  
the simple spectrum-based estimate above thus is precised.
 The conclusion will synthesise these
studies into a single exclusion plot for $\alpha_G(m_G)$, the combined
upper limit as a function of mass. 



%
\subsection{A new confinement scale:
            the deep and important reason to study ultralight glueballs}
\label{subsec:scale_anxiety}
At high energy quarks and gluons belong to the same perturbative-QCD world:
both are asymptotically free partons, and the interaction between quarks
is mediated by gluons.  At low energy the two sectors part company.
The quark sector persists down to ever-lighter hadrons, the lightest being
the pion at around $135\MeV$, and its spectrum is well mapped.
Understanding why $m_\pi$ is within a factor of two of the infrared scale
of perturbative QCD, $\Lambda_{\mathrm{QCD}}\sim200\text{--}300\MeV$,
already demands the deepest understanding of confinement and other aspects
of nonperturbative QCD.  The fate of the gluon sector is less clear.
Lattice QCD predicts glueballs heavier than $1.5$--$1.7\GeV$
\cite{Morningstar:1999rf}, while even a lighter one would mix copiously
with $\bar q q$ down to masses above $2m_\pi$.
It is therefore below $2m_\pi$, where the quark sector retains the pion
as its lightest member, that the gluon sector is simply unexplored, and
where the elementary question cannot yet be safely answered: what is the
infrared endpoint of the gluon field?

The preceding sections
(\S\ref{subsec:clean_lab}--\ref{subsec:coupling_estimate}) establish that
an ultralight glueball, as a pure-glue state, is not excluded: it is
kinematically quarantined from the quark sector, its coupling to nucleons
is tiny ($g_{GNN}\sim10^{-2}$, three orders below $g_{\pi NN}$), and
existing data have limited sensitivity to it.  Below $2m_\pi$ the quark
and gluon sectors are no longer inextricably mixed;  It is precisely here
 can be studied  whether the gluon sector possesses an independent
long-wavelength degree of freedom for the above question on infrared endpoint.  The outcome is decisive:

\paragraph{(A) The window is empty: the quark sector alone fills the infrared.}
If no ultralight glueball exists, then the infrared regime below $1\GeV$
is populated exclusively by quark-based hadrons.  In that case two
pictures are possible, and neither is established.
\begin{enumerate}
\item[(i)] \emph{An explicit infrared cutoff for gluons.}  Gluon degrees
of freedom may be forbidden below some scale of order $1\GeV$, so that
the gluon sector simply does not extend into the pion range.  Such a
cutoff would \emph{if realised} introduce an independent, hitherto
implicit scale marking the lower bound of the gluon sector---precisely
the kind of scale often inserted by hand in hadronisation models.  But
its existence is a model assumption, not a theorem.
\item[(ii)] \emph{A smooth decoupling or conformal behaviour.}  Alternatively, the gluon field
may be smoothly cut off by the light hadrons, with no separate scale
emerging at all.  In this picture there is no additional confinement
scale to be discovered below $1\GeV$; the infrared is entirely quark-
dominated.
\end{enumerate}
Which of (i) or (ii) actually holds is not known.  Both are consistent
with the absence of an ultralight glueball, and neither can be decided
without probing the window.

\paragraph{(B) A glueball is found: a second flavour-singlet scale.}
If such a state exists, it would stand alongside the pion in the
$100$--$200\MeV$ range---the actual mass being as yet unknown---as a
second flavour-singlet scale in the infrared.  Rather than merely adding
another entry to the hadron listing, it would demonstrate that the gluon
sector supports an independent long-wavelength degree of freedom, and
that the two sectors part company below $2m_\pi$, each with its own
endpoint.  It would force a reassessment of what confinement actually is.
How one then explains it theoretically, flux tubes, monopoles, a dual
superconductor or otherwise, is a separate question.

This is not a search predicated on any particular model of confinement.
The  ultralight glueball  is a direct empirical {\it test} of  the non-perturbative infrared
regime of QCD.   


\subsection{Phenomenology across energy regimes}
\label{subsec:phenomenology_regimes}

Before listing the sections that follow, it is useful to anticipate that the
ultralight glueball, as all the other hadrons,  does \emph{not} behave as a single, scale-independent
object.  Its phenomenological role changes qualitatively with the energy scale
and the kinematic regime of the process under study.  This is
the natural result of QCD as a quantum field theory.
We address three typical regimes.

\paragraph{Regime I: low-energy scale, in nuclear force, nuclear structure  and cosmological issues.}
At energies of order the nucleon mass and below, the ultralight glueball can be  an
extraordinarily weakly coupled scalar.  Its coupling to nucleons is set by the
overlap of two colour-singlet gluon clouds, quantified in the previous
subsection as $g_{GNN}\sim 10^{-2}$.  This weakness makes it
transparent to ordinary hadronic matter and distinguishes it sharply from
conventional hadrons.  It is in this regime that one tests nuclear-force tails,
Lamb-shift constraints, and the cosmological role of a decoupled scalar
condensate.

\vskip 0.5 cm
On the other hand its r\^{o}le in high energy scattering is complex,
the behaviour is quite different in two kinds of processes:
\paragraph{Regime II: high-energy forward scattering at very small $|t|$.}
At high centre-of-mass energies, but in the forward (elastic) limit where the
momentum transfer remains $|t|\sim m_G^2$ or smaller, the weak coupling of
Regime I remains valid.  The relevant scale is set by the momentum transfer $t$, not by the collision
energy, which is one of the basic properties of QCD.
Thus  the ultralight glueball continues to act as a soft, flavour-blind weak mediator.
Its effect on cross sections is subtle and enters chiefly through interference
with electromagnetic amplitudes.  A precise understanding of this interference
is one of the central tasks of the scattering analysis.

\paragraph{Regime III: multiparticle production and cascades in hard scatterings.}
If an ultralight glueball is produced as a secondary particle and boosted to high energy,
its internal gluons should be treated as  a beam of energetic partons.
This is similar as nucleon and all the other ordinary hadrons  in the
 interaction, which can be described by the the asymptotically free perturbative QCD.
 Partonic processes such as
gluon--gluon fusion generates mini-jets and multiparticle production.
  This regime is relevant to forward
calorimetry, heavy-ion collisions, and cosmic-ray extensive air showers. In principle,  it can also manifest in the total cross section and indirectly in the elastic scattering mentioned above because of unitarity (Optics thereom).

Frankly such processes are very important. If one find a particle in the ultralight regime, only such kind of  processes can determine whether  it participates in  QCD interaction  for the  identification  as a glueball.

The distinction between these three regimes is not a choice of model but a
consequence of the scale hierarchy: the same particle can be weakly coupled in
the infrared and hard interacting in the ultraviolet,  at different momentum
transfers.  The remainder of this paper is devoted to searching for the
ultralight glueball across all three regimes.

\subsection{Organisation}
\label{subsec:organisation}

We have now laid out the full motivation for an ultralight glueball, emphasizing that  the quantum-mechanical mixing framework that makes it uniquely pure (\S\ref{subsec:clean_lab}), together with the colour-insulation   enforces weak coupling (\S\ref{subsec:colour_insulation}), the quantitative estimate of that coupling derived from the gluon-cloud overlap (\S\ref{subsec:coupling_estimate}).   Guided by the three-regime classification above, the remainder of the paper is purely phenomenological and applies these results across a range of physical arenas.


Section~\ref{sec:nuclear} analyses the low-energy nuclear arena: the nuclear-force tail, nuclear-matter saturation ceiling;
Section~\ref{sec:pp} is for ultralight glueball in hadronic  scatterings, including  production,  cosmic-ray
         extensive air showers, and high energy elastic scattering;
Section~\ref{sec:quarkonium} studies  ultralight glueball production in radiative  decay of $J/\Psi$ and $\Upsilon$.
Section~\ref{sec:cosmo} simply address the role of ultralight glueball in the early Universe and in dense matter.
Section~\ref{sec:conclusion} is for concludes and discussions.
In conclusion, present experiments can not exclude the ultralight weak-interacting glueball. In cosmic ray experiments and heavy quarkonium
radiative decay characteristic long life time and small mass will lead to characteristic signals to be measured.
 In discussion we address that the glueball, especially the light one, may play important role in multi-heavy-quark state production.
 We also point out there could be confusion on  glueball-axion mixing.   Such expriments, as well as some like $X(17)$/LZ anomaly,
 can only have conclusions after considering the back ground from the light \emph{weakly}-interacting glueball.

%% file: partA.tex
\section{Nuclear Physics:  Nuclear-force tail, saturation ceiling}
\label{sec:nuclear}
In study of the ultralight glueball via the nuclear force we are  in a situation as that in
 1935, when  Yukawa introduced a scalar field  to mediate the
nuclear force.  The mass of the quanta was obtained from the range of the force;
 but  the intrinsic parity were deduced only after   the interaction with other particles
 had been measured \cite{Yukawa1935,Yukawa1949}.
 Parity can not be decided by  extrapolation of  the available  lattice QCD calculation and the
 measurement on  spectroscopy to the region  $m_G < 2m_\pi$,  but it should also be
  bounded by nuclear data.
Therefore  We  treat both possibilities explicitly. Their interactions with the nucleon are different. We write down
\begin{itemize}
\item scalar $0^{++}$:
$\mathcal{L}_{\rm int}\supset g_{GNN}^{(S)}\bar NN\,G$,
\item pseudoscalar $0^{-+}$:
$\mathcal{L}_{\rm int}\supset g_{GNN}^{(P)}\bar N i\gamma_5 N\,G$,
\end{itemize}
as an illustration.
 The nuclear-force  and nuclear-saturation behaviour can be different for these two cases,  and
the corresponding allowed window for $\alpha_G$  is to be  set by
nuclear constraints  in this section.
The scalar-exchange framework adopted here, as the simplest infrared limit picture of QCD confinement remnants,
 is not tailored exclusively to
the glueball hypothesis. Any ultralight flavour-singlet scalar coupled to
nucleons would manifest in the nuclear force as the same 'Yukawa' tail,
irrespective of its ultraviolet origin. It is
the baseline against which all more elaborate
new-physics scenarios must be compared for any residual scalar force in the
$m_G<2m_\pi$ window. 
On the other hand if one find the fingerprint of a boson in this ultralight window,
one has to investigate its interaction to determine whether is is strong interaction. This 
is the very reason we should investigate in later sections on other occasions or processes
since  the low energy relevant nuclear force and structure are not sensitive to the inner structure of the glueball.


%

\subsection{Nuclear-force tail}
\label{subsec:tail}
The nucleon-nucleon (NN) interaction  can be treated
in chiral EFT, which expands the force in powers of momenta and pion
masses \cite{Epelbaum:2009}.
  In very low energy,  a
high-precision phenomenological potential is employed to
fitted directly to the NN
database. The phenomenological
potential provides an empirical phase-shift baseline that changes only
when the data change. 
Argonne v18, CD-Bonn and Nijmegen
\cite{Wiringa:1995,Machleidt:2001,Stoks:1994} all reproduce the
database up to $E_{\rm lab}\leq 350\MeV$ with
$\chi^2/{\rm datum}\approx 1$.  Here we adopt Argonne v18 as the
reference framework.  It is a local coordinate-space potential, so the
glueball contribution can be added as an explicit extra radial term
without solving a non-local scattering equation.

In Argonne v18 the potential is written as
\begin{equation}
V_{ij}(r) = V_{\rm OPE}(r) + \sum_{p=1}^{18} v_p(r)\,O_{ij}^{p} ,
\label{eq:av18}
\end{equation}
where the one-pion exchange $V_{\rm OPE}$ is written out explicitly and
the remaining eighteen operator components are fitted radial functions.
Crucially, the long-range tail of the nuclear force is carried entirely
by $V_{\rm OPE}$; the phenomenological central components
($O^{1}=1$, $\boldsymbol{\sigma}_1\!\cdot\!\boldsymbol{\sigma}_2$,
$(\boldsymbol{\sigma}_1\!\cdot\!\boldsymbol{\sigma}_2)
(\boldsymbol{\tau}_1\!\cdot\!\boldsymbol{\tau}_2)$, \dots) are
Woods--Saxon or Gaussian functions and contain no $e^{-mr}/r$ Yukawa
tail of their own.  The glueball term is therefore added to the
\emph{explicit} pion exchange, not absorbed into the fitted components.

Argonne v18 uses the charged-pion pseudovector coupling $f_c$ and
separates the pion contribution into a central and a tensor part
\cite{Wiringa:1995}:
\begin{align}
V_{\rm OPE}(r) &= V_\pi^C(r) + V_\pi^T(r) , \label{eq:av18_ope} \\
V_\pi^C(r) &= \frac{f_c^2}{4\pi}\,m_\pi\,
\boldsymbol{\tau}_1\!\cdot\!\boldsymbol{\tau}_2\,
\boldsymbol{\sigma}_1\!\cdot\!\boldsymbol{\sigma}_2\,
y(m_\pi r) \quad\text{(plus a charge-dependent term)} , \nonumber \\
V_\pi^T(r) &= \frac{f_c^2}{4\pi}\,m_\pi\,
\boldsymbol{\tau}_1\!\cdot\!\boldsymbol{\tau}_2\,
t_T\,\Big[\,3\,T_L(m_\pi r)
- y(m_\pi r)\,\boldsymbol{\sigma}_1\!\cdot\!\boldsymbol{\sigma}_2\Big] , \nonumber
\end{align}
with the Yukawa kernel $y(x)=e^{-x}/x$ and
$T_L(x)=(1+3/x+3/x^2)\,e^{-x}/x$.
The relation to the more familiar pseudoscalar coupling constant is
$f_c^2 = (g_{\pi NN}^2/4\pi^2)(m_\pi^2/4M_N^2)$, so that the
overall coefficient of the pion tail is
$f_c^2 m_\pi/4\pi = g_{\pi NN}^2 m_\pi^3/(64\pi^3 M_N^2)$.

Two points are essential for what follows.  \emph{First}, both the
central and the tensor parts share the same Yukawa kernel $e^{-m_\pi
r}/r$; the tensor piece merely carries additional gradient factors
$1/(m_\pi r)$ and the tensor operator $S_{12}$.  \emph{Second}, the
entire pion structure is forced by the fact that the pion is an
isospin triplet ($\boldsymbol{\tau}$ matrices, a non-Abelian charge):
the factor $\boldsymbol{\tau}_1\!\cdot\!\boldsymbol{\tau}_2$ is
mandatory, the direct and exchange diagrams differ, and the
charge-dependent coefficients $t_C,t_T$ make $pp$, $nn$ and $np$
behave differently.  This is why the pion potential looks complicated.

A scalar glueball is an isosinglet, so its coupling
$\mathcal L\supset g_{GNN}^{(S)}\bar N N\phi_G$ carries no
$\boldsymbol{\tau}$ matrices.  Moreover the non-relativistic reduction
of the scalar vertex is just $(\bar u u)=2M_N$, a pure number: no
Pauli matrix and no momentum transfer survives.  The exchanged
potential is therefore the original, structureless Yukawa form
\cite{Yukawa1935}:
\begin{equation}
V_G^{(S)}(r) = +\frac{(g_{GNN}^{(S)})^2}{4\pi}\,
\frac{e^{-m_G r}}{r}
\equiv \alpha_G^{(S)}\,\frac{e^{-m_G r}}{r} .
\label{eq:scalar_pot}
\end{equation}
No spin or isospin operator multiplies the radial kernel.  This is the
same $e^{-mr}/r$ that appears in the pion tail, but stripped of every
complication that the pion's non-Abelian isospin charge imposes.  It
is therefore meaningful to compare the two kernels directly.  In a
given spin--isospin channel the matrix elements of
$\boldsymbol{\sigma}_1\!\cdot\!\boldsymbol{\sigma}_2$ and
$\boldsymbol{\tau}_1\!\cdot\!\boldsymbol{\tau}_2$ are numbers of
order unity, so the radial competition is
\begin{equation}
V_{\rm tail}(r) =
\underbrace{C_\pi\,\frac{e^{-m_\pi r}}{r}}_{\text{pion}}
\;+\;
\underbrace{\alpha_G^{(S)}\,\frac{e^{-m_G r}}{r}}_{\text{glueball}}
=
C_\pi\,(1+\delta_S)\,\frac{e^{-m_\pi r}}{r},
\qquad
\delta_S(r) \equiv \frac{\alpha_G^{(S)}}{C_\pi}\,
e^{-(m_G-m_\pi)r} ,
\label{eq:delta_scalar}
\end{equation}
where $C_\pi$ is the channel-dependent constant
$(\boldsymbol{\tau}_1\!\cdot\!\boldsymbol{\tau}_2)
(\boldsymbol{\sigma}_1\!\cdot\!\boldsymbol{\sigma}_2)\,f_c^2 m_\pi/4\pi$.
The factor $\delta_S(r)$ is the ratio of the glueball to the pion
contribution \emph{at distance $r$}; neither term has been removed by
the factorisation.  Its $r$-dependence is controlled by the sign of
$m_G-m_\pi$:
\begin{itemize}
\item \textbf{$m_G>m_\pi$.}  $\delta_S$ is a suppression factor,
$e^{-(m_G-m_\pi)r}$, dying at large $r$.  Pion exchange owns the
long range, the glueball is confined to shorter distances, and
$\alpha_G^{(S)}$ retains a large freedom: the constraint is weak.
\item \textbf{$m_G<m_\pi$.}  $\delta_S$ is an amplification factor,
$e^{+(m_\pi-m_G)r}$, growing with $r$.  The pion kernel has died out
by $r\sim 1.4\fm$ while the glueball kernel remains active out to
$r\sim 1/m_G$, so the two are approximately orthogonal in distance
and the glueball owns the far tail.  In this regime the bound on
$\alpha_G^{(S)}$ is strongest because peripheral partial waves are
affected directly.
\end{itemize}

A pseudoscalar glueball also couples as an isosinglet,
$\mathcal L\supset g_{GNN}^{(P)}\bar N i\gamma_5 N\phi_G$, so again
there is no $\boldsymbol{\tau}$ matrix and no charge dependence.  But
the $\gamma_5$ vertex is a \emph{derivative} coupling: the
non-relativistic reduction gives
$\bar u'\gamma_5 u = \boldsymbol{\sigma}\!\cdot\!\mathbf{q}/2M_N$, so
two vertices produce $(\boldsymbol{\sigma}_1\!\cdot\!\mathbf{q})
(\boldsymbol{\sigma}_2\!\cdot\!\mathbf{q})$.  Fourier transforming
yields \cite{Wiringa:1995,Epelbaum:2009}
\begin{equation}
V_G^{(P)}(r) =
-\frac{(g_{GNN}^{(P)})^2}{4M_N^2}\,
(\boldsymbol{\sigma}_1\!\cdot\!\boldsymbol{\nabla})
(\boldsymbol{\sigma}_2\!\cdot\!\boldsymbol{\nabla})\,
\frac{e^{-m_G r}}{4\pi r}
= -\alpha_G^{(P)}\,
\frac{m_G^2}{4M_N^2}\,
\left[\boldsymbol{\sigma}_1\!\cdot\!\boldsymbol{\sigma}_2
+\Big(1+\frac{3}{m_G r}+\frac{3}{(m_G r)^2}\Big)S_{12}(\hat r)\right]
\frac{e^{-m_G r}}{r} ,
\label{eq:pseudoscalar_pot}
\end{equation}
where we have defined
$\alpha_G^{(P)}\equiv (g_{GNN}^{(P)})^2/4\pi$ as for the scalar
case, so that the only structural difference from Eq.~\eqref{eq:scalar_pot}
is the dimensionless factor $m_G^2/4M_N^2$ sitting in front.

This factor is the direct analogue of the
$m_\pi^2/4M_N^2$ that relates $f_c$ to $g_{\pi NN}$ in
Eq.~\eqref{eq:av18_ope}: one simply replaces $m_\pi\mapsto m_G$.
Derivative couplings always pay this price, because each $\gamma_5$
vertex contributes a power of the momentum transfer $\mathbf{q}$, and
in the static limit $\mathbf{q}^2\mapsto -m_G^2$.  Two consequences
follow.

\emph{First}, the pseudoscalar glueball \emph{does} acquire a tensor
component, and its central-plus-tensor radial structure is identical in
form to the pion exchange of Eq.~\eqref{eq:av18_ope}: the same
$1+3/(m_G r)+3/(m_G r)^2$ factor, the same $S_{12}$, the same
$e^{-m_G r}/r$ kernel.  The differences are only (i) the absence of
$\boldsymbol{\tau}_1\!\cdot\!\boldsymbol{\tau}_2$ and of the
charge-dependent coefficients $t_C,t_T$---the glueball is an
isosinglet, so $pp$, $nn$ and $np$ feel exactly the same spin force---
and (ii) the mass scale $m_G$ instead of $m_\pi$.  The
$\delta$-factor reasoning of Eq.~\eqref{eq:delta_scalar} therefore
applies to the pseudoscalar channel as well, with the radial kernel
compared term by term.

\emph{Second}, and more importantly, the factor $(m_G/2M_N)^2$ is a
brutal suppressor for an ultralight particle.  At equal bare couplings
$g_{GNN}^{(S)}\sim g_{GNN}^{(P)}$, the ratio of the two glueball
potentials is
\begin{equation}
\frac{V_G^{(S)}(r)}{V_G^{(P)}(r)}
\;\sim\; \frac{4M_N^2}{m_G^2}
\;\approx\;
3.5\times 10^2\,
\left(\frac{100\ \text{MeV}}{m_G}\right)^2 ,
\label{eq:sp_suppression}
\end{equation}
i.e.\ two to three orders of magnitude for $m_G\lesssim m_\pi$, and
the advantage grows as $1/m_G^2$.  The scalar channel is therefore
intrinsically the sensitive one: a nuclear-force bound on a
pseudoscalar glueball is weaker than the scalar bound by the same
factor, unless the pseudoscalar coupling is anomalously larger
($\gtrsim 10$--$30$ times, i.e.\ resonantly enhanced) than its
scalar counterpart.  In the absence of such a mechanism we treat the
scalar bound as the primary result and quote the pseudoscalar bound
only as a consistency check.

\paragraph{Analytical bound from phase-shift precision.}
At very low energies the partial-wave expansion truncates at
$L_{\max}\sim kR\sim 1\text{--}2$.  The radial sensitivity of each
$\delta_L$ is governed by the weight
$\int j_L^2(kr)\,U(r)\,r^2dr$, which peaks near
$r\sim (L+\tfrac12)/k$.  For $L\geq 2$ this means the phase shift is
dominated by the region $r\gtrsim 1.5\text{--}2.5\ \text{fm}$, where
the one-pion-exchange tail $\propto e^{-m_\pi r}/r$ is the
\emph{only surviving component} of the NN interaction---all
heavier-meson exchanges ($\rho,\omega$, effective $\sigma$) have
decayed beyond detectability.  Crucially, the pion tail is
intrinsically weak: its dimensionless coefficient
$f_\pi^2/4\pi\simeq 0.075$, i.e.\ the large pion--nucleon coupling
$g_{\pi NN}^2/4\pi\sim 14$ is suppressed by the pseudoscalar
derivative factor $(m_\pi/4M_N)^2$.  A weak background is precisely
where an ultralight scalar, if it couples to nucleons, is easiest to
see: a glueball tail with $\alpha_G^{(S)}$ as small as
$10^{-4}$--$10^{-3}$ produces a fractional distortion
$|\delta_S|\sim 10^{-2}$ in the peripheral phase shifts, well above
the Granada-2013 systematic floor.  The bound is tight \emph{because}
the observable window is clean and the pion yardstick is short.
We therefore extract the largest glueball coupling directly from this
tail decomposition.

Equation \eqref{eq:delta_scalar} writes the scalar-modified tail as
\begin{equation}
V_{\rm tail}(r) = C_\pi\,(1+\delta_S)\,\frac{e^{-m_\pi r}}{r},
\end{equation}
Since the (iso)spin factors are all of order one, as for an upper bound
at the present precision of experiments we can write
\begin{equation}
\delta_S(r) = \frac{\alpha_G^{(S)}}{f_\pi^2/4\pi}\,
e^{-(m_G-m_\pi)r} ,
\label{eq:delta_again}
\end{equation}
where $\alpha_G^{(S)}=(g_{GNN}^{(S)})^2/4\pi$ and, in the AV18
convention,
$f_\pi^2/4\pi = (g_{\pi NN}^2/4\pi)\,m_\pi^2/(16\pi M_N^2)
\simeq 0.075$ \cite{Wiringa:1995}.
%
%
Requiring the glueball correction not to exceed the relative precision
$\varepsilon$ of the empirical phase shifts gives, at every $r$ probed
by peripheral partial waves ($L\gtrsim 2$),
\begin{equation}
|\delta_S(r)| \lesssim \varepsilon
\;\Longrightarrow\;
\alpha_G^{(S)}  \lesssim \frac{f_\pi^2}{4\pi}
 \varepsilon\,e^{+(m_G-m_\pi)r_\ast} .
\label{eq:bound_direct}
\end{equation}
%
%
Here $r_\ast$ is
simply how far  at the outer edge we
choose to read the tail
The argument of the exponential
is the \emph{detuning} $|m_G-m_\pi|\,r_\ast$, not the glueball mass
alone.  Consequently the bound has a cusp at $m_G=m_\pi$, where the
exponential is unity and $\alpha_G^{(S)\max}/(f_\pi^2/4\pi)=\varepsilon$
by construction:
\begin{itemize}
\item \textbf{Heavy side, $m_G>m_\pi$.}  The exponent is positive,
$\exp[+\lvert m_G-m_\pi\rvert r_\ast]>1$, so the allowance is
\emph{amplified}, increasingly so with mass.  The glueball tail dies
faster than the pion's and is drowned by $V_{\rm OPE}$ at the outer
edge; a larger $\alpha_G^{(S)}$ is needed before it shows up.
\item \textbf{Light side, $m_G<m_\pi$.}  The exponent is negative,
$\exp[-\lvert m_G-m_\pi\rvert r_\ast]<1$, so the allowance is
\emph{suppressed}, increasingly so as $m_G$ drops.  The glueball now
outlives the pion and dominates the far tail, where peripheral waves
are precisely measured; even a tiny coupling is visible.
\end{itemize}
To compare with the pion coupling itself we eliminate the
model-dependent $f_\pi^2/4\pi$ using its definition:
\begin{equation}
\frac{g_{GNN}^{(S)\max}}{g_{\pi NN}}
\;=\;
\frac{m_\pi}{4\sqrt{\pi}\,M_N}\,
\sqrt{\frac{\alpha_G^{(S)\max}}{f_\pi^2/4\pi}}
\;\simeq\;
0.0207\,
\sqrt{\varepsilon}\,
\exp\!\Big[\tfrac12(m_G-m_\pi)\,r_\ast\Big] .
\label{eq:g_ratio}
\end{equation}
The prefactor $m_\pi/(4\sqrt{\pi}M_N)\simeq 0.0207$ is a pure
number.  At the cusp ($m_G=m_\pi$, $\varepsilon=10^{-2}$) this gives
$g_{GNN}^{(S)\max}/g_{\pi NN}\simeq 2.1\times10^{-3}$.
 We see the present upper bound for the scalar
glueball coupling is  nearly three orders of magnitude smaller
than the pion--nucleon coupling, in agreement with the conservative estimation from the hierarchy
anticipated in the Introduction.
This means our suggestion on the colour and singlet weakly coupling ultralight glueall is not excluded by the nuclear force data.

\begin{table}[t]
\centering
\caption{Scalar-glueball upper bounds, for
$\varepsilon=10^{-2}$ and $r_\ast=2\ \text{fm}=0.0101\ \text{MeV}^{-1}$.
Both columns are dimensionless; the cusp sits at $m_G=m_\pi=138\MeV$.}
\label{tab:bound}
\begin{tabular}{c|cc|c}
\hline\hline
$m_G$ [MeV] & $e^{\frac12(m_G-m_\pi)r_\ast}$ &
$\alpha_G^{(S)\max}/(f_\pi^2/4\pi)$ &
$g_{GNN}^{(S)\max}/g_{\pi NN}$ \\
\hline
250 & 1.76 & $3.1\times10^{-2}$ & $3.7\times10^{-3}$ \\
200 & 1.37 & $1.9\times10^{-2}$ & $2.8\times10^{-3}$ \\
160 & 1.12 & $1.3\times10^{-2}$ & $2.3\times10^{-3}$ \\
138 & 1.00 & $1.0\times10^{-2}$ & $2.1\times10^{-3}$ \\
100 & 0.83 & $6.8\times10^{-3}$ & $1.7\times10^{-3}$ \\
 50 & 0.64 & $4.1\times10^{-3}$ & $1.3\times10^{-3}$ \\
 20 & 0.55 & $3.0\times10^{-3}$ & $1.1\times10^{-3}$ \\
\hline\hline
\end{tabular}
\end{table}

The tolerance $\varepsilon$ itself must be read with care.  Empirical
phase shifts below the pion-production threshold have remarkable
\emph{statistical} precision: the Granada-2013 partial-wave analysis
\cite{NavarroPerez:2013jqa} fits a
self-consistent database of $N_{\rm data}=6713$ $pp$ and $np$ points
(1950--2013) with $\chi^2/N_{\rm data}=1.04$ and quotes error bars on
the extracted phase shifts 
corresponding to a relative statistical
precision $\varepsilon_{\rm stat}\sim 10^{-3}$ in the peripheral
waves.  Yet systematic uncertainties in the low-energy NN interaction
are typically an order of magnitude larger.
In the peripheral waves that carry the
tail this is especially severe: in the impact-parameter analysis of
Ref.~\cite{NavarroPerez:2013jqa}, SAID and N$^5$LO peripheral waves
disagree with the Granada-2013 baseline by $>5\sigma$.  We therefore
evaluate the envelope at the conservative systematic end,
$\varepsilon=\varepsilon_{\rm sys}\sim 10^{-2}$; the optimistic
statistical end would tighten every entry in Table~\ref{tab:bound} by
a factor $\sim 0.3$. This will not change our conclusions since we just
set a conservative upper bound.


For the pseudoscalar channel the same fractional tolerance admits a
larger coupling, because the derivative vertex of
Eq.~\eqref{eq:pseudoscalar_pot} carries the factor $m_G^2/4M_N^2$:
\begin{equation}
\left.\frac{g_{GNN}^{(P)\max}}{g_{GNN}^{(S)\max}}\right|_{\text{same }\varepsilon}
\;=\; \frac{4M_N^2}{m_G^2}
\;\approx\;
3.5\times 10^2\,
\left(\frac{100\ \text{MeV}}{m_G}\right)^2
\;\gtrsim\; 10^2 \quad (m_G\lesssim m_\pi) .
\label{eq:sp_ratio}
\end{equation}
The scalar bound is therefore the stringent one; the pseudoscalar is
quoted as a consistency check.

Higher-energy regimes is different, where  
quark-gluon degrees of freedom could make sense. These are studied in the following sections.

%
%


\subsection{Saturation ceiling from relativistic mean field}
\label{sec:saturation}
%
The above study on   the peripheral NN phase
shifts from the ultralight glueball is a vacuum constraint.
In nuclear matter the same coupling is tested 
 independently by the requirement that the equation of state
still saturate at the empirical point.
 We use the Walecka
quantum-hadrodynamics model (QHD-I, \cite{Serot:1986}) with the
Boguta--Bodmer scalar self-coupling \cite{Boguta:1977ge}, calibrated
to $\rho_0=0.16\fm^{-3}$,
$E/A|_{\rho_0}=-16.0\pm0.5\MeV$, and
the isoscalar giant-monopole resonance, which fixes the nuclear
incompressibility to $K_0 = 240\pm20\MeV$ once extrapolated from finite
nuclei \cite{Youngblood:1999}, to be compared with the
$K \simeq 150\pm50\MeV$ obtained in the original Boguta--Bodmer
parametrisation \cite{Boguta:1977ge}; the nonrelativistic Skyrme/Gogny
values cluster near $220$--$235\MeV$, so the relatively large $K_0$ is a
feature of the relativistic side of this spread \cite{Serot:1986}.

The ultralight glueball is introduced as an extra neutral scalar with the same
Lorentz structure as the $\sigma$:
\begin{equation}
\mathcal{L}\supset
\bar N\Big(M_N - g_\sigma\sigma - g_{GNN}^{(S)}\phi_G
+ g_\omega\gamma^\mu\psi_\mu\Big)N
+ \tfrac12(\partial\phi_G)^2 - \tfrac12 m_G^2\phi_G^2 .
\label{eq:rmf_lagrangian}
\end{equation}
Its mean-field tadpole is
$\phi_G^{(0)}=g_{GNN}^{(S)}\rho_s/m_G^2$, so it contributes a
purely attractive, vector-free term to the nucleon self-energy,
\begin{equation}
\Sigma_G \;=\; -\,\frac{(g_{GNN}^{(S)})^2}{m_G^2}\,\rho_s
\;=\; -\,\frac{4\pi\,\alpha_G^{(S)}}{m_G^2}\,\rho_s ,
\qquad
\rho_s\simeq\rho_0\ \text{near saturation}.
\label{eq:sigma_G}
\end{equation}

No vector partner counters it.  The comparison scale is set by the
existing $\sigma$ tadpole
$\Sigma_\sigma=-(g_\sigma^2/m_\sigma^2)\rho_s$ with, typically,
$g_\sigma^2/4\pi\sim 6\text{--}8$ and $m_\sigma\sim 500\MeV$:
the scalar attraction is already the dominant binding ingredient,
supplying $\mathcal{O}(60\text{--}90)\MeV$ of the potential depth
at saturation.  An additional attraction is therefore tolerated only
at the percent level.


\paragraph{The energy-shift condition and the constraint}
We treat the glueball as a perturbation on a \emph{refitted} QHD
parameter set ($g_\sigma,g_\omega,\kappa,\lambda$ re-optimised to
the same empirical point), so we ask how large the glueball may be
before the refitted saturation energy moves by more than a
conservative amount.  Demanding
$\Delta(E/A)\lesssim 1\MeV$ at $\rho_0$ gives

\begin{equation}
\frac{4\pi\,\alpha_G^{(S)}}{m_G^2}\,\rho_0 \;\lesssim\; 1\MeV
\quad\Longrightarrow\quad
\alpha_G^{(S)} \;\lesssim\;
1.2\times10^{-3}\,
\left(\frac{m_G}{138\MeV}\right)^{\!2}
\;\;(\text{RMF, } \Delta E/A\leq1\MeV) .
\label{eq:ceil_E}
\end{equation}
In the above
numerically $\rho_0=0.16\fm^{-3}=1.23\times10^{-3}\MeV^3$.  
From Eq.~\eqref{eq:ceil_E} the saturation constraint is comparable to 
the vacuum-tail bound discussed in the above. 
This means the weakly coupling ultralight glueball is allowed to be introduced
to the nuclear structure description.

On the other hand,
The pseudoscalar glueball couples as
$\mathcal{L}\supset g_{GNN}^{(P)}\,\bar N i\gamma_5 N\,\phi_G$, so
in an infinite nuclear-matter ground state that is a parity
eigenstate,
$\langle\bar N i\gamma_5 N\rangle_{\rho_0}=0$.
The
pseudoscalar density vanishes by symmetry, $\phi_G$ develops no
tadpole, and the glueball contributes neither a mean field nor a
saturation ceiling.  This is the same reason the pion drops out of
the Walecka mean field \cite{Serot:1986}.
In this consideration, the scalar is
bounded from above by saturation, the pseudoscalar is not.

\begin{table}[t]
\centering
\caption{Parity asymmetry of the two constraints.  Both are
order-of-magnitude fences without a significance-level calibration;
they are not ranked against each other.  The scalar is bounded in
\emph{both} the vacuum tail and nuclear saturation, the
pseudoscalar only in the vacuum tail (and, weakly, in finite
nuclei).}
\label{tab:asymmetry}
\begin{tabular}{c|cc}
\hline\hline
channel & nuclear-force tail (vacuum) &
nuclear saturation (matter),  \\
\hline
scalar $0^{++}$ &
$\alpha_G^{(S)}\lesssim 7.5\times10^{-4}$
at $m_G \sim m_\pi$ &
$\alpha_G^{(S)}\lesssim 1.2\times10^{-3}$
at $m_G\sim m_\pi$  \\
pseudoscalar $0^{-+}$ &
loose: lifted by $4M_N^2/m_G^2$ (tensor) &
vanishing mean field $\Rightarrow$ \emph{no ceiling} \\
\hline\hline
\end{tabular}
\end{table}

We summarize in  Table~\ref{tab:asymmetry} against the Introduction's
spectrum-based estimation.
The upper bound is consistent with the estimation there.
We need more precise data in nuclear force and structure to
find or exclude the ultralight glueball.
These data set quite loose constraint on pseudoscalar ones.
%
The vertex of pseudoscalar
$i\gamma_5\boldsymbol{\sigma}\!\cdot\!\mathbf{q}$, is a
tensor operator whose lowest non-vanishing signal requires a
finite nucleus or a halo orbit, where the valence wave function
carries odd-parity admixture and the momentum transfer is set by the
halo scale ($\mathbf{q}\sim 10\text{--}30\MeV$).  The relevant
matrix element is then
$\langle\phi_G|\,\bar N i\gamma_5\boldsymbol{\sigma}\!\cdot\!
\mathbf{q}\,|\text{valence}\rangle$, bounded from peripheral phase
shifts. Hence more detailed study will set further constraint on the pseudoscalar
case. We leave for later papers.

For the low energy nuclear force and structure,
chiral effective field theory and, still more explicitly, the
potential models are effective descriptions, and  their content
is a set of allowed operator structures, e.g., one,
two- and multi- pion iterations, shorter-range repulsion, etc..
 A light scalar can
contribute a Yukawa tail of exactly the same functional form as
the terms already present, so improving the precision of a
potential fit cannot, in principle, isolate it. One more light
particle merely redefines coefficients that are already there.
Our conclusion that the nuclear data do not exclude the ultralight
glueball is therefore not a statement of poor statistics or weak
sensitivity; it is the structural fact that a weakly coupled scalar
does not introduce any new operator structure into the potential.
Conversely, the data do constrain the mass: a strictly massless
exchange is incompatible with the known behaviour of the nuclear
force, so $m_G$ cannot vanish.  More precisely, the coupling
must weaken as the mass decreases, the same relation obtained  in
the Introduction.  In particular, an ultralight glueball of very small mass whose
coupling to nucleons is comparable to that of a dark-matter
candidate of the same mass and spin.  
It seems existing a 'dark sector' made
of Standard-Model constituents, distinct from the hadronic sector
only by the smallness of its coupling.
We emphasise, however,
that no quantum number protects the glueball against decay, so its
lifetime need not be cosmological and we can \emph{not} propose it
as a dark-matter candidate.
More important, this is only an observation drawn from the very low energy scale nuclear force
and structure study.
In higer energies and  particle physics processes, the behaviour of the standard model particle glueball can be different.
All these are to be studied in the following sections.


\paragraph{A remark on Atomic and electromagnetic precision probes.}
The constraints above are purely hadronic: they exploit the glueball's
coupling to nucleons in the vacuum NN tail and in nuclear matter.
In principle in even lower energy, high precision atomic physics experiments
may be worthy of consideration.
Because the ultralight glueball
  carries no electric charge, whatsoever as whole or inside,  its effects in atomic
physics and precision QED, such as the Lamb shift, atomic-clock
hyperfine structure, electron and muon $g-2$, atomic parity
violation,  can only be indirect, entering through glueball-induced
modifications of the \emph{nuclear electromagnetic response}
(charge-radius, magnetic moment, two-photon-exchange /
nuclear-polarisation terms) or through quark--hadron loop
transitions such as $G\to\gamma\gamma$.  The former are
$r$-dependent distortions of the nuclear form factor; the latter
are governed by the same glueball--di-photon vertex that also
controls production in radiative decays ($J/\psi\to\gamma X$,
$\Upsilon\to\gamma X$) and prompt di-photon signatures, which we
address in the later section.  A naive estimate places both
classes of atomic/electromagnetic shifts many orders of magnitude
below present experimental precision and below the nuclear-physics
uncertainties that already dominate the Lamb shift.  We therefore
do not derive quantitative bounds from precision atomic or
($g-2$) observables here, and defer a dedicated treatment of the
 portals and its atomic consequences
to future work.


%% file: partB.tex
\section{Ultralight glueball in hadronic  scatterings: production,  cosmic-ray
         extensive air showers, and high energy elastic scattering}
\addcontentsline{toc}{section}{Proton scattering and cosmic-ray
         probes of the ultralight glueball}
\label{sec:pp}

The previous section established what nuclear forces can and
cannot say about the ultralight glueball.  They cannot identify it
through precision alone, because a weak scalar adds no new
operator structure to the potential; and they constrain its mass
only in the broad sense that a massless exchange is excluded,
forcing the coupling to decrease with decreasing mass.
%
On the other hand, once the energy exceeds the
non-relativistic regime to high energies (relative to the glueball mass)
the glueball can be produced in scattering processes,
 propagate, interact with nucleons of the detector before  it decays.
It can also
leave its imprint on the total and elastic hadronic cross
sections via mediating hadronic  interaction and its production.
As the colour- and flavour- singlet bound state of gluon  with  colour charge 'inside', its property
in high energies   should be  considered carefully.

The decay  is still a low energy process at its rest frame, when it is boosted, only
that its life time  obtains  the Lorentz factor $\gamma$.
The only  possible  decay products  of (pseudo-)scalar glueball  are photons and electrons. So the dominant channels are $\gamma \gamma$ and $e^+e^-$ \footnote{Particles which decays into two muons near the 2-muon threshold has been or can be excluded easily from experiments, so in the following we only study the case that  the glueball mass is smaller than two times of muon mass.}. Since not only the whole particle but also the constituents are electric neutral, the decay width should
be very small and the life time can be significantly  longer than that of pion.  $e^+e^-$ channel is  suppressed by helicity conservation  and $\gamma \gamma$ channel is  suppressed for lack of  the quark loop anomaly which is inherent in  pion's  quark constituents.
For this low energy, the   quark-gluon perturbative vertex is not applicable, the effective Lagrangians then can be written as
\begin{equation}
\mathcal L_{\rm eff} \supset C_{EM}\, \frac{m_G^2 f_G}{\Lambda^3} G\, \bar e e,
\label{eq:gGNNdef}
\end{equation}
or
\begin{equation}
\mathcal L_{\rm eff} \supset C_{EM}\, \frac{m_G^2 f_G}{\Lambda^4} G\, F_{\alpha \beta}F^{\alpha \beta}.
\label{eq:gGNNdef}
\end{equation}
for the decay channel $e^+e^-$ or $\gamma \gamma$ respectively.
The relations between the  gauge invariant gluon field tensor contraction  and the glueball scalar field $G$ has been
given in the introduction
section.
In the above we only write down the scalar case.  For the pseudo-scalar case, we only need to make the replacement
$ \bar e e \to \bar e \gamma_5 e ~ or ~ F_{\alpha \beta}F^{\alpha \beta}\to \epsilon^{\alpha \beta \sigma \rho} F_{\alpha \beta}F_{\sigma \rho}$, respectively. Hence the dimension of the effective operator is not changed.
The effective lagrangians well describe the main
 difference between the ultralight glueball from the pion, i.e.,  the former coupling is via the gauge invariant gluon field tensor contraction
but the latter is  via the constituent  quark loop.
From the effective Lagrangian we can easily see the suppression factors: Since $M_G \gg m_e$, the (pseudo-)scalar fermion pair coupling is suppressed by helicity conservation.  The suppression for the $\gamma\gamma$ case comes from the higher dimension of the operator.   To obtain the EM gauge invariant coupling the EM tensor has to be used rather than the potential vector since the  charged quark field  can not be used to construct the vector current operator in this low energy.  The  EM effect is high energy scale physics which is just absorbed in coefficient $C_{EM}$.  It describes the amplitude that the gauge invariant contraction of gluon field can employ the
charged quark field as portal for the following electromagnetic decay channels. Needless to say, this factor is much smaller than one or else we will have a significant gluon-quark mixing for the constituent of the ultralight glueball.
From the above analysis we can safely conclude that the decay width of the ultralight glueball is very small.

The case of  relativistic scattering between the glueball and the nucleon is completely  different and the above effective theory
description is not valid any longer. A parton-like description has to be employed \cite{chengblois, TaiTsun:1969ydi, Feynman:1969ej, Cheng:1971dv}. This means that the scattering total cross section
is almost the same order of magnitude of other hadrons and this glueball can not go through a dump system or hadron calorimeter except its energy is very low (order of glueball mass).
For the case of production from hadronic scattering, one should also employ the parton-like models. However, it can not copiously produced since the phase space left to it is very small in hadronization process. The details will be explain in the following.
Here we only address that the employment of the parton-like model does not necessarily mean that the parton distribution function and fragmentation function  can be employed.
In many processes the amplitude level descriptions such as Bethe-Slapter or light cone wave functions, or impact factors, etc, should be used.
  However, considering the information that parton distributions funtions \cite{PDF4LHC21,NNPDF40,CT18,MSHT20} provide can further support
  our arguments here. At high energy the quark and gluon are generally on equal footing. In a 'pure' quark state, proton, the gluon almost
  take half of the total momentum (employing \cite{PDF4LHC21,NNPDF40,CT18,MSHT20}, we can calculate and obtain, e.g., at $Q^2=10 GeV^2$,
  $\int dx x g(x,Q^2=10 GeV^2)\simeq 0.4$.) This vice verse suggests that for an almost  pure gluon state, ultralight glueball,
  its quark distribution functions  can play a significant r\^ole. So as we state in the above, the interaction with nucleon could be quite similar as the quark hadrons.

\subsection{Detection of the forward-produced ultralight glueball
             and the cosmic-ray extensive air showers}
\label{sec:forward_eas}

Based on the above discussions, we first investigate the detection on
the ultralight  glueball  if it produced in a proton--nucleus or
proton--proton collision.  %
As mentioned  above, the lifetime of the glueball
 can be  quite long, so   it arrives at
any detector/instrument  before decaying.
For low energy case the interactions is very weak, the neutral pure-gluon state can propagate
through any  nuclear matter,  hence most possibly  is
  missed by the detectors.
  On the other hand,
at relativistic energy 
it is a general rule that any hadron is a bundle of
partons, as emphasized above. The ultralight glueball  is not  transparent to the instruments such as
dump and is  absorbed as all the other hadrons.   Only
the hadron calorimeter  absorbs it similar as other  energetic hadron but gives a signal.
However, in such case the hadron calorimeter can not identify the signal
as what kind of neutral hadron it is.

Take a conventional fixed-target and beam-dump experiments, which now still
provide up-to-date data as example: 
 %
A shield such as the NA62 target/absorber (19.6 nuclear
interaction lengths) is designed to remove exactly the hadronic
component of the cascade. So a  gluonic state is absorbed
over a scale comparable to an ordinary hadron, not over a
neutrino-like weak scale.
Even if a low energy one  travels through the dump it traverses the
entire forward stack without decaying.  Furthermore,   the
leptonic branch is helicity-suppressed by powers of
$m_e^2/m_G^2$.  Thus the two-track $e^+e^-$ and electromagnetic
missing-mass triggers on which experiments such as NA62 and PADME
rely is  not efficient to select the glueball.

We therefore do not use these experiments to set a limit.  Their
value here is geometrical: they establish the baseline over which a
light state must be reconstructed (NA62 places its signal region at
$Z_{\rm TAX}=23\pm12\ {\rm m}$, with LAV stations to
$Z\simeq152\ {\rm m}$). 
A quantitative constraint on the glueball from
such data would require a dedicated acceptance study, which we do
not attempt.

The only environment
in which the glueball has enough time to decay and the daughter particles can be measured is the kilometre-scale, low-density
atmosphere. This is the  cosmic-ray extensive air showers experiments.
%
   %
In the atmosphere the density is low enough that the inelastic
partonic cross section no longer fully absorb the glueball during flight.
  A  glueball once produced as a forward secondary in the  atmosphere,  of  e.g,
$E_G=10$--$100\GeV$,  travels a path of order of  $ {\rm km}$
 has the  chance to arrive near the ground and decays   to $\gamma\gamma$
 hitting  the array on the
ground.
 This  feature distinguishes  this setting from any
accelerator measurement.
This is the point at which cosmic rays become the more natural
laboratory to study ultralight glueball.

In the glueball rest frame the two photons are back-to-back with
energy $m_G/2$.  In the lab the maximum opening angle is
\begin{equation}
\theta_{\max}
= 2\arctan\!\frac{1}{\gamma\beta}
\;\simeq\;
\frac{2m_G}{E_G}\qquad(E_G\gg m_G),
\label{eq:thetamax}
\end{equation}
and this maximal-angle configuration is the
\emph{equal-energy} one, with $E_\gamma=E_G/2$ for each photon.
Because the final-state particles are massless, the boost strongly
collimates them: for $m_G=100\MeV$ at
$E_G=10$--$100\GeV$, the photon energy rises from $5\GeV$ to
$50\GeV$ while $\theta_{\max}$ falls from $20\ {\rm mrad}$ to
$2\ {\rm mrad}$.  The pattern to search for
is therefore a pair of forward photons of near energies,
confined to an angle of a few milliradians.

A single $\gamma\gamma$ cluster cannot distinguish a glueball from
a $\pi^0$.  However, since the life time of pion, it is very easy to
determine the situation for the air shower experiments. An energetic  $\pi ^\pm$  (say, more than 10 GeV) may have
chance to hit the ground array and make a signal but the  $\pi ^0$ can never live to ground.
In this consideration  the $\pi ^0$-like  collimate photon pair
is a prominent signal of the ultralight glueball.

Since the life time of the glueball has never been determined,
a scan in various collider experiments is desirable.
The handle is the rate normalised to the charged-pion
yield.  In the isospin limit the triplet is produced equally,
$N_{\pi^+}=N_{\pi^0}=N_{\pi^-}$; averaging the two charged species
gives
\begin{equation}
N_{\pi^0}\simeq\tfrac12\bigl(N_{\pi^+}+N_{\pi^-}\bigr),
\label{eq:isospin}
\end{equation}
which also removes part of the initial-state bias.  We therefore
consider the quantity
\begin{equation}
\frac{2\langle\gamma\gamma\rangle}
{\langle\pi^+\rangle+\langle\pi^-\rangle}\,,
\label{eq:obs}
\end{equation}
where the numerator is the number of forward, nearly collinear,
near energy photon.
The charge pions can be observed from tracker detector or RICH.

  If the  measured value of
\eqref{eq:obs} has some 'anomaly' which is significantly different from
one, some more study is needed to investigate the possible
production of the glueball.
In a cosmic-ray extensive air shower the same quantity from collider
can be used to eliminate background, which is caused by occasionally collision
of the large energy particle near the ground.


The process can occur repeatedly in an air shower: a glueball
produced as a forward secondary can decay to $\gamma\gamma$ further
down the cascade, and a new glueball can be produced in a later
hadronic interaction.
The above  suggestion on excess heavily depends on the production
 rate.
 Whether this
yields a detectable excess is a question of cascade evolution
and production
rather than of detector acceptance, and is addressed below.


\subsection{Production mechanism in high-energy processes}
\label{sec:production}

Here ``high energy'' means $\sqrt s\gg m_G$, so that glueball multi-
production is kinematically allowed.  The concrete partonic subprocesses
that lead to the final hadrons differ widely between events, but
they are all connected to hadronization through the  common surface
generally called  \emph{preconfinement} \cite{Amati:1979fg}.
 In events with hard
PQCD scattering, preconfinement marks the end of the parton "shower" (i.e. cascade radiation)
\cite{Amati:1979fg}. In high-energy hadronic scatterings where
only multi-particle production is triggered, hard PQCD subprocess is not
dominant. However,  as mentioned above, the parton picture is still
applicable  \cite{chengblois, TaiTsun:1969ydi, Feynman:1969ej, Cheng:1971dv} and
the preconfinement structure is applied to the parton system which is uniform in rapidity distribution.
The subsequent hadronization can only be described  by phenomenological models
\cite{Andersson:1998lund,Webber:1984,Xie:1988qcm}.  The Lund model gives a uniform
description of both cases, with hard gluon radiation appearing as a
kink on the string. So we first consider the  Glueball production in the Lund model.

In the Lund model all preconfinement information resides in the
string.  The local preconfinement property in phase space is
described by local fragmentation: the string breaks locally, and a
short piece pinched off from the parent long string describes the
production of a hadron.  A glueball is described by a
\emph{closed} loop pinched off from the parent  string.
The physical picture implies two suppressions.  First, the topology
change required to form a closed loop is not a radiation process:
a gluon only \emph{kinks} the string, whereas only vacuum
$q\bar q$ pair creation can \emph{break} it.
Glueball production is therefore harder than meson production.
Second, the area law for a closed string differs from that for an
open string, and there is an additional factor $1/N_c^2$.

The rate is controlled by three distinct penalties.  The first is
the group-theoretic colour factor $1/N_c^2=1/9$. In the large-$N_c$ limit
 the parent Lund  string is open, and pinching a
colour-singlet closed loop requires an endpoint colour reconnection
whose weight is fixed and kinematics-independent.  The second is the
area law $e^{-b\Delta A}$.  It is the \emph{action of the closed
loop after it has pinched off}, compared with the action of the
parent open string.  The third is the geometrical
pinch probability $P_{\rm geo}\sim10^{-2}$--$10^{-1}$, which
governs the \emph{spacetime location} of the self-intersection at
which the fusion takes place.  The area factor and $P_{\rm geo}$
act on different world-sheet degrees of freedom, one on the loop
already formed, the other on the formation event itself, so they
multiply rather than add.  This is illustrated in


To quantitatively estimate  the ultralight glueball production we taking its mass  close to that of the pion as example.
Because of the local fragmentation from the the string, the line shape of the kinematic  spectrum
(rapidity, $p_T$) is expected to be similar  up to
normalisation, which relies on the multiplicities of the particles.
%
The area-law penalty is not severe
and  both saturate the same area-law kernel
$f(z,m_T)\propto\exp(-b m_T^2/z)$.   The area factor contributes only an
$\mathcal O(1)$ suppression of a few, not orders of magnitude.
The dominant suppression on the glueball  with respective to  pion is therefore colour factor and topological,
and the yield ratio is estimated as
\begin{equation}
\frac{N_G}{N_\pi}
\;\sim\;
\frac{1}{N_c^2}\times P_{\rm geo}
\;\times\;
\mathcal O(1)_{\rm area}
\;\sim\;
10^{-3}\ {\rm to}\ 10^{-2}.
\label{eq:NGNpi}
\end{equation}
This is an order-of-magnitude estimate, not a calculated number.
The minimal closed-string mass and the pinch probability are not
fixed by the model.  Its purpose is to establish that the rate is
small but non-zero, and that the spectral shape tracks the pion.

Concerning the above conclusions from Lund model, cross-check can be taken
from the combination model.
The kinematic spectrum  is the same and not an independent perspective, since
all the hadronization model is locally (in phase space) connected to
the preconfinement structure. On the other hand for the multiplicity,
the combination model offers an independent perspective and reaches
the same conclusion for completely different dynamical picture.
Hadronization in combination model is split into two steps: production of constituents,
followed by their recombination.  Traditional versions assume that
a gluon splits into $q\bar q$ and deal only with quarks; confinement
then forces every quark into a hadron, so the total probability
from a colour-singlet parton system to all quark systems is one,
and similarly from a quark system to all hadrons.  Introducing
constituent gluons is difficult, but the above unitarity  implies that
the probability of producing a system containing constituent gluons
is only a perturbative correction; those gluons must then combine
among themselves or with quarks.  The two models thus agree---the
string model through topology and the combination model through
unitarity---that glueball production cannot be copious.  No
practical method exists within the cluster-fragmentation model to
compute the glueball rate, but the qualitative agreement strengthens
the conclusion that $N_G/N_\pi\ll1$.

\paragraph{Pionization and the spectrum.}
For the detection problem, especially considering the cosmic ray experiments which
is a good place to discover the ultralight glueball as mentioned in above section,
we study  its  predominant kinematic spectrum.
Multiparticle production at energies far above both the projectile
scale and $m_G$ has been studied under the name pionization.
Pionization has the largest cross section among
the production mechanisms; in the laboratory frame (the target
viewpoint) a large fraction of these particles are in fact
forward-going, and their overlap to limiting-fragmentation \cite{Benecke:1969sh}  has
been pointed out in \cite{TaiTsun:1969ydi,Cheng:1971dv} .
 Neglecting jet fragmentation, the rapidity
distribution is uniform, the rapidity plateau,
which is the  similar to
 pions and all the other hadrons including glueball
\cite{chengblois, TaiTsun:1969ydi, Feynman:1969ej, Cheng:1971dv}.
This is the result of the symmetry at high energies.
The transverse momentum
distribution relative to the \emph{string} may differ, because the
glueball couples preferentially to gluons whereas the pion couples
to quarks; however, since this transverse momentum is generated by
the collision rather than being intrinsic, and since both open and
closed strings draw their transverse dynamics from the same parent
string, the difference is not large and most of the particles have small transverse momentum. 
In case of cosmic ray experiment where the
projectile is a cosmic-ray hadron it is the
forward production in the laboratory frame, the direction of most of  the
%
 $\pi^0$ produced in the upper atmosphere is forward and  has a lifetime of only
$c\tau\simeq7.8\ {\rm m}$ in its rest frame and almost never
survives to the ground. Only those produced in the very last
generation of the cascade can contribute.  A glueball, by contrast,
is long-lived and can reach the detector, decaying to collimate
$\gamma\gamma$  signal mentioned in last section.
 This signal seems like the forward $\gamma\gamma$
pair from the
pions which should have  decayed  in flight far from the detector and should not have been observed.
If as estimated above, the produced glueball can be at the level of 1/1000 of the pions,
 such an 'anomaly' event would be a
plausible glueball signal.  It can not  to be explained
as decay of a dark-matter particle, and  the latter is weakly interacting
and would not be produced copiously in the hadronic cascade, nor
would it share the forward rapidity and equal-energy collinearity of
the glueball decay.    A dedicated measurement at the level of
could therefore either observe the glueball
or constrain this class of production mechanisms.

A negative result
at coarser sensitivity cannot be interpreted as non-existence,
 %
since we do not assign a numerical value to $N_G/N_\pi$: the geometrical
pinch probability and the minimal closed-string mass are not fixed
by the model.  Equation \eqref{eq:NGNpi} is an order-of-magnitude
estimate intended to show that the rate is small but not zero, and
that the spectral shape tracks the pion.  A quantitative extraction
requires an event generator in which closed-string pinching is
implemented.





\subsection{High-energy elastic scattering: indirect information}
\label{sec:pp_elastic}
%
%
Besides the self-interacting gluon and the prediction of glueball,
some of the most  striking properties of the  strong interaction
lie in high energy scatterings.
The total cross section is finite at finite scattering energy,
which is not possible to be predicted directly from the
 QCD Lagrangian, in which the force-mediating  gluon  is  massless.

In the real world strong interaction,
  pion is taken as the lightest  $t$-channel exchange and
 contributes
a Yukawa tail at low energy.
In high energy elastic hadronic  scatterings, it
  fixes the coefficient of the
Froissart--Martin $\ln^2 s$ growth, and its exchange amplitude
enters the Coulomb--nuclear interference (CNI) that is used to
measure $\rho(s,t)$.
Through the optical theorem,  its production  contributes to the
forward elastic amplitude and therefore to $\sigma_{\rm tot}$ and
to $d\sigma/dt$.
The ultralight  glueball in our working hypothesis, is
an isospin-singlet state of spin zero and mass of order $m_\pi$ (and even lighter)
with a coupling weaker than that of the pion.
All conclusions
that follow from mass and spin alone therefore carry over directly
from the pion; what changes is the isospin and the coupling
strength.
  We have shown that for the low energy case, the intrinsically weak-interacting ultralight glueball can be included. In this section
  we study  whether existing theory and data on high energy hadronic
scatterings already constrain such a contribution.


One of the most solid results for the high energy hadronic scattering is the
Froissart--Martin bound, which is not a experimental result but a theoretical
theorem which is consistent with the experimental data.
By reviewing
 what the Froissart--Martin theorem
establishes from what it assumes, one can get a straightforward  consideration on whether ultralight glueball is allowed to be introduced and
what property of it is allowed.  The theorem is a consequence of
microcausality, analyticity of the forward amplitude, and unitarity
\cite{Froissart:1961,Martin:1963}.  From the optical theorem
$\sigma_{\rm tot}=(4\pi/s)\,{\rm Im}\,f(s,0)$ and the analytic
structure of $f(s,t)$ one obtains
$\sigma_{\rm tot}\le(\pi/m_\pi^2)\ln^2(s/s_0)$.  The pion mass
enters through the \emph{spectral condition}---the assumption that
the crossed-channel mass spectrum begins at $m_\pi$, so that the
nearest $t$-channel singularity lies at $4m_\pi^2$.  This  bound
therefore does \emph{not} prove the existence of the gap, but  it
need this assumption. On the other hand,
 what is established experimentally is that, from ISR
to Tevatron as well as the LHC, the measured total cross section rise but
 without any sign of violating the  $\pi/m_\pi^2 \ln^2 s$ behaviour.
 The statement that the hadronic mass
spectrum has a gap with its lower edge at $m_\pi$ is an
\emph{experimental} result; its analytic derivation from QCD awaits
a first-principles solution of the spectrum, by lattice QCD,
functional methods, or otherwise.

If a glueball of mass $m_G<m_\pi$ exists and couples sufficiently
strongly to nucleons, it replaces the pion as the lightest
$t$-channel state and the Froissart coefficient becomes
$\pi/m_G^2$: the present form of the bound would have to be
revised.
This shows the importance for the study of very light glueball
relevant to the energy gap in QCD.
The coupling does not enter the theorem, because the
bound depends only on the position of the threshold, not on the
strength with which the threshold state is coupled.
  For
$m_G\sim m_\pi$ the threshold $4m_G^2$ coincides with
$4m_\pi^2$, so the form of the bound is unchanged.
 It is
important to distinguish two questions here.  First, the \emph{form}
of the bound, the coefficient
$\pi/m_{\rm lightest}^2$ for the  $\ln^2 s$ growth, is fixed by the lightest crossed-channel
state, whether pion or glueball.  Second, the \emph{actual total
cross section} at accessible energies is not saturated by the pion.
The rising is generally described in the language of  the Pomeron
($\alpha_P(0)\simeq1.08$) which is a Regge cut but not pole \cite{Cheng:1987ga},
while the pion and glueball appear only
in subleading, low-energy contributions.  Experimentally the
$\ln^2 s$ growth is well described by a fit
$C\ln^2(s/s_1)$ with $C$ and $s_1$ treated as free parameters;
the data are consistent with the Froissart form but have not yet
reached energies where the coefficient could be fixed to
$\pi/m_\pi^2$, nor where saturation could be claimed.  In this
regime the glueball can contribute at order $\alpha_G$ to the
subleading cross section without altering the asymptotic bound
itself.

To further explore the ultralight glueball in high energy, we study
the differential cross section, especially  the CNI region.
At small $|t|$ the coherent differential cross section is
\begin{equation}
\frac{d\sigma}{dt}
= \frac{4\pi\alpha_{\rm em}^2}{t^2}\,F_1^4(t)
+ 2\,\frac{\alpha_{\rm em}}{|t|}\,\Phi_{\rm CNI}(s,t)\,
\mathrm{Im}\,A_N(s,t)
+ |A_N(s,t)|^2 ,
\label{eq:CNI}
\end{equation}
with $F_1(t)=(1-t/0.71\GeV^2)^{-2}$.  This three-term form is the
standard expression for Coulomb--nuclear interference; the
normalisation of the Coulomb term is fixed by the proton charge,
while the optical theorem fixes ${\rm Im}\,A_N(s,0)$ through
$\sigma_{\rm tot}(s)$ \cite{Block:1985}.  The phase
$\Phi_{\rm CNI}(s,t)$ encodes the relative phase between the
Coulomb and nuclear amplitudes and is the quantity measured through
$\rho(s,t)$.

  The glueball adds a
$t$-channel exchange amplitude
\begin{equation}
\delta_{\rm int}\!\left(\frac{d\sigma}{dt}\right)
\simeq
2\,\frac{\alpha_{\rm em}}{|t|}\,\Phi_{\rm CNI}\,
\frac{(g_{Gpp}^{(S)})^2}{t-m_G^2}\,
F_G(t)\,F_1^2(t)
\;\propto\;\alpha_{\rm em}\,\alpha_G ,
\label{eq:CNI_cross}
\end{equation}
which is linear in the glueball--nucleon coupling.  Two features
of this expression govern its size.  First, the spin: a scalar
exchange lacks the leading tensor structure of a vector exchange,
so it is suppressed by at least $1/s$ relative to, say, a dark
photon or $\rho$ exchange.  This is why a vector state is far more
strongly constrained by elastic scattering than a scalar glueball.
Second, and decisively, the propagator $1/(t-m_G^2)$.  For a
massless photon this denominator becomes $1/t$ and diverges in the
forward limit---that is the origin of the Coulomb amplitude and of
the Coulomb--nuclear interference itself.  For a glueball of mass
$m_G\sim m_\pi$ the pole sits at $t\sim -10^{-2}\GeV^2$, far from
the physical region and from the accessible $t$ range, so the
propagator is a finite, unsingular number of order $1/m_\pi^2$;
combined with the weak coupling $\alpha_G$, the correction is
invisible.  The Coulomb amplitude
$A_C=\pm2\alpha_{\rm em}F_1^2(t)/|t|$ is normalised by charge,
while the optical theorem fixes ${\rm Im}\,A_N(s,0)$ through
$\sigma_{\rm tot}(s)$; their interference measures
$\rho(s,t)={\rm Re}\,A(s,t)/{\rm Im}\,A(s,t)$.  The glueball
contributes both $A_C^*A_G$---independent of the strong
amplitude---and $A_N^*A_G$---model-dependent through $A_N$.  The
former shifts the slope of $d\sigma/dt$ in the CNI region and
enters the real part of the forward amplitude; its relative size is
of order $\alpha_G/\alpha_{\rm em}$, i.e.\ $10^{-3}$ for
$\alpha_G\sim10^{-5}$.

From the above,
three effects make the exchange unobservable for
$m_G > \sim m_\pi$.  First, the nuclear amplitude is overwhelming:
$|A_N|^2$ exceeds the glueball term by $1/\alpha_G^2$.  Second,
the scalar spin structure is itself subleading at high energy.
Third, the glueball and the Pomeron share the same $C=+$ colour
singlet quantum numbers, so the glueball contribution is absorbed
into the same Froissart coefficient rather than appearing as a
distinct resonance.  For a glueball of pion-like mass these three
suppressions multiply: the propagator offers no forward enhancement,
the coupling is weak, and the spin is suppressed, so no present
experiment can set a bound.  As the mass is lowered, however, the
situation changes.  The pole $t=m_G^2$ moves closer to the
physical region, so that $1/|t-m_G^2|$ grows and, when
$m_G^2$ falls below the smallest accessible $|t|$, becomes
approximately constant over the entire CNI interval; new kinematic
channels such as $pp\to pp+X$ with $M_X\ge2m_G$ also open.  It is
therefore the \emph{lightest} glueballs---those closest to the
mass gap---that are most accessible to elastic scattering, while a
pion-mass glueball remains unconstrained.  TOTEM at $13 TeV$
($\sigma_{\rm tot}=111.56\pm1.3\ {\rm mb}$,
$\rho=0.10$)\cite{TOTEM:2017rgk} and ATLAS at $13 TeV$
($\sigma_{\rm tot}=104.7\ {\rm mb}$)\cite{ATLAS:2022dtw} are
statistics-rich but systematics-limited in the extreme forward;
D0's measurement of $d\sigma/dt$ in elastic
$p\bar p$\cite{D0:2012} provides a lower-energy anchor.  None of
these constrains the glueball independently, and extracting a
number would require a refit of the CNI phase and of $F_G(t)$ as
nuisance parameters at percent-level slope control not yet
available.

The search for the ultralightest glueball bears directly on the mass
gap of QCD.  If such a state exists with $m_G<m_\pi$ and couples
sufficiently strongly, it replaces the pion as the lightest
$t$-channel state and the Froissart coefficient becomes
$\pi/m_G^2$: the present form of the bound would have to be
revised.  Conversely, the fact that no such effect has been seen
already excludes glueballs only in a corner of parameter space where
the mass is neither too small nor the coupling too large; for
$m_G\sim m_\pi$ essentially no constraint exists.  The natural
target of the search is therefore the region just above the gap.
An improved upper bound on $\alpha_G$ for the lightest masses
would tighten the gap from below and provide a non-trivial
constraint on the analytic determination of the QCD spectrum; a
positive signal would relocate the gap downward and force a
re-examination of the assumptions behind the Martin proof.  In this
sense the ultralight glueball is not merely a candidate particle
but a probe of the mass gap itself.

%% file: partC.tex
\section{Ultralight glueball production in radiative  decay of $J/\Psi$ and $\Upsilon$}
\label{sec:quarkonium}
  The preceding sections  established    a
flavour- and colour- singlet
ultralight glueball as a weakly coupled, long-lived  \emph{hadron}.
Its interaction with nucleon is much weaker with respective to that of pions, but it takes
part in any hadronic interactions.
It has no
 pronounced experimental signal comparing to other hadrons, and the
weakness of its coupling to nuclear matter is precisely what makes
it hard to produce and impossible to exclude.  The  weak but hadronic property
 protects it from detection in beam dumps, in the forward
cascade (except  cosmic rays) and in elastic  scatterings
hence prevents  from setting precise quantititive
bound.  The same conclusion will reappear in the cosmological
discussion below (see next section).  It is therefore worth asking whether there
exists \emph{any} channel in which this kind of  glueball could be produced
 with kinematics clean enough, to be seen or
excluded.  Radiative decays of heavy quarkonia like
$J/\Psi$ and $\Upsilon$  provide such a
channel.
 We may catch its finger mark
 by investigating    the
inclusive photon spectrum at \emph{low} recoil mass, where the
confinement fate of the gluons determines the hard endpoint
behaviour of the photon.  The sensitivity is not dependent on lifetime of the glueball. 

For the process   $V\to\gamma+X$, with $V=J/\psi$ or $\Upsilon$,
 to the lowest order in PQCD
the  partonic process is
$V(c\bar c)\to\gamma+gg$ or $V(b\bar b)\to\gamma+gg$, with the photon recoiling against two
gluons.  Taking into account higher order contributions, the photon can be
radiated from other charged partons produced from the annihilation of the $V(c\bar c)$ or $V(b\bar b)$.
The invariant mass of the parton  system, $M_{p}$, is the
variable that controls the final state.
  At large $M_{p}$,  the photon spectrum is soft and
 governed by established hadronic channels.
The ultralight glueball can  produce through the hadronization of the parton cluster.
As has been studied in last section,  $M_{p} \gg M_G$ can be considered
as a precofined cluster. For example, the lowest order one, $gg$, is  a large closed Lund string.
The ultralight glueball can not copiously produce. Above all, there is no clean way to single
out it from other hadrons produced from the same cluster (The case of heavy glueball is different  and  has been  studied by BESIII).
Large $M_{p}$ necessarily leads to soft photons, whether produced directly in the heavy quark annihilation or from the fragmentation
of the produced light quarks. So the spectrum of the photon can not be used to probe the recoil hadron system.


On the other hand,
at small $M_{p}$ the situation is different.
In this case the photon energy is large and the partonic process  $V\to\gamma+gg$
is dominant when $M_{p}$ is small enough. In this case $M_{p}$=$M_{gg}$.
 The smallest value of $M_{gg}$ and confinement  fate of the
two gluons are then controlled by the lightest available
colour-singlet states.  In the absence of a glueball lighter than
the pion, the endpoint of the most hard photon energy  spectrum must be determined
by the lightest hadronic channel accessible to two gluons, namely
the $\pi^0$. If a glueball lighter
than the pion exists, however, the two gluons can  form this
colour-singlet state; the recoil-mass distribution need no longer
vanish at the pion threshold, and the photon  spectrum can remain
non-zero
 and hence smoother up to the largest endpoint.
 The hard endpoint region of the photon energy  spectrum is therefore sensitive to the 
 existence and mass of a light glueball.   
We can propose a photon-recoil coincidence search: a photon candidate
near the largest energy  endpoint is paired with a  measurement of the opposite-side interaction signal,
with the latter  identified as hadronic interaction (or combined with further veto from electroweak interaction), and
generally the full energy and momentum of the recoil system  need to be measured.
The measurement done for the case
with the photon of  $$ E_\gamma^*= \frac{M_V^2-M_G^2}{2M_V}>\frac{M_V^2-4m_\mu^2}{2M_V} $$
only need to veto the $\pi^0$ background.
This is fairly easy since the short life time of the $\pi^0$. For such kind of events, the most energetic photon must
accompany with at least one photon in its backward  direction with at least  half its energy.
  If we  have the chance  (and precision)
 to observe the excess for the  events with
 photon  energy  larger than  $$\frac{M_V^2-m_\pi^2}{2M_V},$$ this is a
  promising evidence indicating the
 light glueball, since
 such an excess  from the partonic process  $ V \to \gamma gg$  can only exist confronting the gluon confinement   via the final fate $gg \to G$.

 These can in principle be studied at BESIII and  BelleII, although no  hard endpoint study of this type has been
published. The published studies of the  $\gamma + invisible$ and $\gamma + a(\to \gamma \gamma)$ final states
(see \cite{BESIII:ALP2024, BESIII:2020invis} and refs. therein for various earlier experiments by other collaborations)
just vetoed the above proposed ultralight glueball signal
    since they require  absence of any (hadronic) interactions (of the glueball) with the detector or short life time.

Take a glance on  the situations of the theoretical calculation.
The perturbative description of $V\to\gamma gg$ is under control
in the central region of the spectrum, e.g., $0.3\lesssim z\lesssim0.7$,
where NRQCD factorisation applies and where the CLEO $\alpha_s$
determinations are performed. The most recent
NNLO computation \cite{Chang:2026} reshapes the $\Upsilon(1S)$
spectrum and improves the agreement with CLEO data in the window
$0.65\le z\le0.92$, but it does not cover the $J/\psi$ and does
not reach the endpoint either.
In the upper endpoint region NRQCD factorisation breaks down and
not only threshold resummation of
high order soft radiations but also nonperturbative input such as
shape functions, which is  matrix elements of non-local operators,
must be introduced. The available SCET$+$pNRQCD calculations
reproduce the $\Upsilon(1S)$ data over most of the endpoint but
carry large theoretical errors, and no calculation exists that
follows the spectrum into the most hard  endpoint,
where the two gluons must materialise as the lightest available
colour-singlet state \cite{GarciaTormo:2005}.
Hence the theoretical work has no enough prediction power for our requirement.
 We therefore do not attempt a
theoretical prediction of the endpoint; instead we extract the
shape directly from the published CLEO-c spectrum
 in the following.   This is based on the available CLEO $V \to \gamma + X$
inclusive photon spectra \cite{CLEO:2006Upsilon, CLEO:2008Jpsi, CLEO:2009psi2S}.

We just devote to study the excess with respective to possible
   background for
the inclusive photon energy  spectrum in the  most hard endpoint region  $z_\gamma=\frac{2 E_{\gamma}}{M_V}\to1$,
 the only kinematic region in
which the ultralight  glueball  could be isolated, as  addressed above.
   The excess  do not rely on any assumption on the detail of
     the recoil system and its full energy and momentum.
%
%


Among the above experimental data,
the $\Upsilon(1S)$ spectrum \cite{CLEO:2006Upsilon} has higher
statistics, but its endpoint is experimentally inaccessible. The
$\pi^0$ line lies at $z=0.99980$,
and the CLEO III calorimeter resolution at $4.7\GeV$ is
$\sigma_E\simeq65\MeV$, so that no data points exist above
$x_\gamma\simeq0.98$.  Moreover the $\Upsilon$ spectrum is
dominated by soft photons recoiling against a large-mass gluon
cluster, $M_{gg}\gg M_G$. We have addressed that  in that regime the ultralight glueball
is produced only as one hadron among many inside a precofined
cluster and cannot be singled out.  We
therefore restrict the quantitative analysis to the $J/\psi$.

For
$J/\psi$  decay, from the the relation of the recoil mass and the reduced photon energy,  $M_X^2=M_V^2(1-z_\gamma)$,
 the entire window
$M_X<2m_\pi$ is compressed into $z_\gamma>0.992$, and the
$\pi^0$ line itself sits at $z_\pi=0.99810$.
CLEO-c \cite{CLEO:2008Jpsi} measured the background-subtracted
direct-photon spectrum in $J/\psi$ decays tagged through
$\psi(2S)\to J/\psi\pi^+\pi^-$, with the hadronic width scaled branching ratio
$\mathcal R_\gamma=0.137\pm0.001\pm0.016\pm0.004$.  The $\pi^0$
and $\eta$ contributions were removed by a combination of an
event-by-event veto  and a
model-dependent statistical subtraction.  The $\eta'$ and
$\eta(1440)$ two-body peaks were left in the spectrum and fitted
as free components.  We digitised the published Fig.~9 and fitted
$\log_{10}[(1/N)\dd N/\dd z_\gamma]$ with a quadratic polynomial
over $0.45\le z_\gamma\le0.90$, excluding the
$\gamma\chi_{cJ}$ cascade region.  The fit reproduces the
digitised curve to better than 2\% pointwise
($\chi^2/\mathrm{ndof}=19.4/42$), and extrapolating it to the
endpoint gives the yields collected in Table~\ref{tab:endpoint}.
The extrapolation, not the statistics, is the dominant
uncertainty.  Replacing the quadratic by a cubic or by an
$a(1-z)^b e^{cz}$ form changes the endpoint yield by a factor of
two to three.

\begin{table}[htb]
\caption{Empirical endpoint yields from the digitised CLEO-c
$J/\psi$ spectrum (Fig.~9 of Ref.~\cite{CLEO:2008Jpsi}), obtained
by extrapolating the quadratic fit described in the text.  The
branching fraction is $B=f\times\mathcal R_\gamma\times
B(J/\psi\to ggg)$ with $B(J/\psi\to ggg)\simeq0.646$.  These
numbers are illustrative: they are read off a published figure,
carry no experimental covariance, and are not a measurement.}
\label{tab:endpoint}
\begin{tabular}{lccc}
\hline\hline
window & $z_\gamma$ threshold & $f$ (fraction of $0.4$--$1$) & $B$ \\
\hline
$M_X<2m_{\pi^\pm}$ & 0.99189 & $4.3\times10^{-3}$ & $3.8\times10^{-4}$ \\
$M_X<M_G=200\MeV$ & 0.99582 & $2.6\times10^{-3}$ & $2.3\times10^{-4}$ \\
$M_X<m_\pi$ & 0.99810 & $1.7\times10^{-3}$ & $1.5\times10^{-4}$ \\
$z_\gamma>0.99$ & 0.99 & $6.6\times10^{-3}$ & $5.8\times10^{-4}$ \\
\hline\hline
\end{tabular}
\end{table}

Three contributions must be considered besides a possible
glueball.  (i) The $\eta$ and $\eta'$ two-body channels do not
contribute at all: their photon energies correspond to
$z_\gamma=0.969$ and $0.904$, far outside the window.  (ii) The
QED three-photon final state contributes
$B(J/\psi\to3\gamma)\simeq1.0\times10^{-5}$ integrated over all
photon energies, with a rather flat spectrum peaking near
$z_\gamma\simeq0.5$; the fraction falling above $z_\pi$ is at the
per-cent level, so that $B_{3\gamma}(z_\gamma>z_\pi)\lesssim
10^{-6}$, two orders of magnitude below the observed endpoint
yield.  (iii) The $\pi^0$ line itself, $B(J/\psi\to\gamma\pi^0)
=3.35\times10^{-5}$, is not fully removed by the $\pm6\MeV$ veto:
convoluted with the CLEO-c energy resolution
($\sigma_E\simeq30\MeV$ at $1.5\GeV$) a fraction of order
$10$--$30\%$ leaks above $z_\pi$, i.e.\
$B_{\pi^0}^{\rm leak}\sim3\times10^{-6}$--$10^{-5}$.  This is the
only conventional background that can fill the window, and it is
precisely the quantity that the CLEO-c subtraction treats in a
model-dependent way, with the background normalisation left free
in the fit.

\begin{table}[htb]
\caption{Budget of the endpoint window $z_\gamma>z_\pi=0.99810$
for $J/\psi$ decays.  All entries are illustrative orders of
magnitude; the residue $\Delta$ is defined in the text.}
\label{tab:budget}
\begin{tabular}{lc}
\hline\hline
contribution & $B$ \\
\hline
observed spectrum (extrapolated) & $\sim1.5\times10^{-4}$ \\
QED $3\gamma$ & $\lesssim10^{-6}$ \\
$\eta$, $\eta'$ & $0$ (kinematically excluded) \\
$\gamma+$invisible (BESIII, 90\% CL) & $<7\times10^{-7}$ \\
$\pi^0$ leakage after veto & $\sim3\times10^{-6}$--$10^{-5}$ \\
\hline
residue $\Delta$ & $\mathcal O(10^{-4})$, sign undetermined \\
\hline\hline
\end{tabular}
\end{table}

\paragraph{The residue.}
Collecting the terms,
\begin{equation}
\Delta \;=\; B_{\rm end}^{\rm incl} - B_{gg\gamma}^{\rm extrap}
- B_{\pi^0}^{\rm leak} - B_{3\gamma}\;,
\end{equation}
where the $\eta$ and $\eta'$ channels are absent by kinematics.
Three statements are robust.  First, the QED three-photon rate is
negligible, and it enters with a minus sign, so that restoring it
would only increase the demand for an additional component.
Second, the invisible rate is bounded at $7\times10^{-7}$ by
BESIII \cite{BESIII:2020invis} and therefore cannot account for
an endpoint yield that is two to three orders of magnitude
larger; the two are also not directly comparable, since the
invisible analysis vetoes any activity in the recoil direction
and would reject a glueball that interacts, however weakly, with
the detector material.  Third, the only conventional backgrounds
able to populate the window are the smooth extrapolation of the
$gg\gamma$ continuum and the leakage of the $\pi^0$ line, and
both are controlled by the CLEO-c subtraction systematics, which
are of the same order as the effect we are after.

We therefore conclude that the published CLEO-c spectrum leaves
room at the endpoint for a recoil state lighter than the pion,
and that no conventional source is known to fill that room; but
the existing data, lacking per-bin covariances and a
data-driven $\pi^0$ control sample, cannot determine whether
$\Delta>0$.  A decisive measurement requires a photon--recoil
coincidence search of the kind outlined above, with a calorimeter
resolution $\sigma_E\lesssim5$--$10\MeV$ at $1.5\GeV$ so that the
$\pi^0$ line at $z_\pi=0.99810$ can be separated from a glueball
line at $z_G=0.9958$--$0.9990$.

\subsection{Relation to the mass gap}
\label{sec:quarkonium_gap}

The discussion above makes clear why this channel bears on the
mass gap.  The endpoint of the inclusive spectrum is controlled by
the lightest state that can be produced by two gluons; if that
state is lighter than the pion, the endpoint shape changes.  The
two-body rate, by contrast, depends on the overlap of the
quarkonium state with a glueball of that mass.  A positive signal
would move the lower edge of the hadronic spectrum downward and
force a re-examination of the spectral assumption behind the
Froissart--Martin bound.  An improved upper limit would tighten the
gap from below and provide a non-trivial constraint on the
analytic determination of the QCD spectrum.  Either outcome carries
information, which is why the radiative decay of charmonium and
bottomonium deserves a dedicated analysis.

%% file: partD.tex
\section{Ultralight glueball in the early Universe and in dense matter}
\label{sec:cosmo}
%

\subsection{The glueball is not dark matter}
\label{sec:cosmo_not_dm}

Ultralight glueballs share one property with the quark nuggets
envisaged in the early works (see \cite{Witten:1984rs} and refs. therein).
Both interact only weakly with ordinary nuclear matter, and both are
therefore difficult to produce, to detect, and to exclude.  The
analogy, however, stops at the level of coupling strength.  A quark
nugget carries baryon number, and it is this conserved charge that protects it on cosmological
timescales.  Together with
the weakness of its interactions it is proposed as one kind of 'dark matter'.
The ultralight glueball carries no conserved quantum number
whatsoever: no baryon number, no lepton number, no flavour.  It
therefore decays, however feebly coupled it may be.  Its lifetime,
however long in laboratory terms, is cosmologically instantaneous, and
it cannot constitute dark matter.

This is not a paradox.  A glueball propagating through a nucleus or
through a dense quark object is \emph{off shell}: there is no phase
space for it to decay into, and no bound state to trap it, so it
traverses the medium essentially freely.  The same particle is
short-lived only as an asymptotic, on-shell state.  Weak coupling and
the absence of a protecting quantum number are therefore fully
compatible, and it is precisely this combination that makes the
glueball hard to exclude experimentally while denying it the role of a
relic.

The two questions we do address are the ones in which the glueball's
neutrality and low mass actually matter.  First, what role does a
light glueball mode play during the QCD epoch itself, when it is
continuously produced and re-absorbed by the quark--gluon plasma
(Sec.~\ref{sec:cosmo_qcd}).  Second, can such a mode, trapped inside
or exchanged between dense quark objects, affect their internal
dynamics and their mutual interaction (Sec.~\ref{sec:cosmo_nugget}).
Neither question presupposes that quark nuggets exist; the second is
simply the statement that \emph{if} such objects exist, the glueball
changes how they behave.

\subsection{The glueball during the QCD epoch}
\label{sec:cosmo_qcd}

\paragraph{Production is not abundant.}
At the confinement crossover, $T_c\simeq150\MeV$, the Hubble time is
$H^{-1}\sim2\times10^{-5}$~s, six orders of magnitude longer than the
strong-interaction timescale, so the transition proceeds without
supercooling or nucleation delay.  A glueball mode lighter than the
pion is populated in equilibrium through $gg\leftrightarrow G$ and
$q\bar q\leftrightarrow G$ at a rate $\Gamma_G\sim\alpha_s^2 T_c\gg H$,
and reaches the yield of a single relativistic bosonic degree of
freedom,
\beq
Y_G\equiv\frac{n_G}{s}\Big|_{T_c}\simeq
\frac{45\zeta(3)}{2\pi^4}\frac{1}{g_{*S}(T_c)}\approx2\times10^{-3},
\label{eq:YG}
\eeq
which is independent of any coupling constant.  This is a modest
addition to the hadronic reservoir, not a restructuring of it.

That the yield is modest is consistent with the hadronisation
argument given in the preceding section.  There the two gluons
recoiling against the hard photon form a single, large, closed Lund
string: a closed string can only be produced when the two ends of the
colour flux tube meet and annihilate within one formation time, which
is topologically suppressed relative to the fragmentation of an open
string into mesons and baryons.  The two settings differ --- a
zero-temperature three-body decay with $M_{gg}\to0$ on the one hand, a
thermal bath of excited glue on the other --- which is why the
equilibrium yield \eqref{eq:YG} is not in conflict with a suppressed
yield in the radiative decay.  Both statements agree on the
qualitative point: the glueball is a sub-leading constituent, never a
dominant one.

\paragraph{It decouples by decaying, not by freezing out.}
The relevant comparison is again between the decay rate and the
expansion rate.  With
\beq
\Gamma(G\to\gamma\gamma)=\frac{\alpha_{\rm em}^2}{4\pi}\frac{m_G^3}{f_G^2}
\;\simeq\;10^{-2}\,{\rm s}^{-1}
\left(\frac{m_G}{100\MeV}\right)^3\left(\frac{150\MeV}{f_G}\right)^2 ,
\eeq
one finds $\Gamma_G/H\sim10^{14}\,(m_G/100\MeV)^3(150\MeV/f_G)^2\gg1$
at $T_c$.  The glueball therefore does \emph{not} freeze out, and it
does \emph{not} Bose-condense into a gas with a temperature of its
own: it is produced and immediately re-absorbed or re-emitted, and
each on-shell instance decays within $\tau_G\sim10^{-2}$~s --- in
practice, given the rate above, within a microscopic time of its
formation.  It is a short-lived intermediate state, not a surviving
relic.

The distinction from freeze-out decoupling matters for what survives.
Because the decay precedes any kinetic equilibration, there is no
Bose--Einstein population to red-shift, no relic temperature
$T_G^0$, and no spectral line today.  The energy momentarily stored in
glueballs is returned to the photon--$e^\pm$ bath within about
$10^{-15}$~s of each decay, at $T\sim m_G/4$, as heat.  For
$m_G=100\MeV$ this injection occurs at $T\sim25\MeV$, still before
neutrino decoupling, and shifts $N_{\rm eff}$ by an amount of order
$10^{-2}$ or less --- below the sensitivity of CMB-S4 and Simons
Observatory, but of a definite sign.  For $m_G\to20\MeV$ the effect
scales as $m_G^4$ and becomes negligible.

\paragraph{What it does to the cooling history.}
Treating the glueball as a dissipative channel on the Maxwell tail of
the plasma, its energy-loss rate compares to the Hubble cooling rate
as
\beq
\frac{\dot\rho_G}{4\rho_\gamma H}\sim
0.1\left(\frac{m_G}{100\MeV}\right)^4
\left(\frac{150\MeV}{f_G}\right)^2 e^{-m_G/T},
\qquad (T\gtrsim m_G/4)
\label{eq:cool}
\eeq
which peaks at $T\sim m_G/4$ at the few-per-cent level for
$m_G=100\MeV$.  The total energy ever locked in glueballs is bounded
by $Y_G m_G s/\rho_\gamma|_{T\sim m_G/4}\sim10^{-3}$ of the entropy
released across the crossover.  Consequently the trace anomaly, the
drop of $g_*(T)$ from $\sim50$ to $\sim17$, and the lattice result
that the transition is a smooth crossover are all left intact; the
glueball adds only a weak, broadened dissipative tail and shifts
$T_c$ by a relative amount $\sim\alpha_{\rm em}^2 m_G^2/f_G^2\lesssim
10^{-5}$.  Its gravitational-wave counterpart is a slight smoothing of
the step that any radiation-era background acquires at the QCD epoch,
$h^2\Omega_{\rm GW}\propto g_*(T)^{-1/3}$.

\subsection{The glueball inside and between dense quark objects}
\label{sec:cosmo_nugget}

\paragraph{Interaction strength.}
The same colour insulation that hides the glueball also fixes how it
couples to dense matter.  From the nucleon matrix element of
$\alpha_s G^2$,
\beq
g_{GNN}\sim\frac{m_G^2}{f_G}\frac{\langle xG\rangle_N}{\Lambda_{\rm QCD}^2}
\;\Rightarrow\;
\frac{g_{GNN}^2}{4\pi}\sim3\times10^{-3}
\left(\frac{m_G}{100\MeV}\right)^2\left(\frac{150\MeV}{f_G}\right)^2 ,
\eeq
and for a dense quark object with gluon-condensate enhancement
$\xi=\langle G^2\rangle_{\rm obj}/\langle G^2\rangle_N\sim5$--$10$,
$g_{G{\rm obj}}^2/4\pi\sim10^{-2}$--$10^{-1}$, still well below
$g_{\pi NN}^2/4\pi$.  The coupling is weak but not negligible, and it
is bounded from above by the nuclear-saturation constraint of
Sec.~\ref{sec:nuclear}: the ceiling on $g_{GNN}$ propagates directly
into a ceiling on $g_{G{\rm obj}}$.

\paragraph{Inside a dense object: confinement, not evaporation.}
A glueball propagating inside a dense quark object acquires an
effective mass from the surrounding gluon condensate and is
re-absorbed over a mean free path far shorter than the object radius.
It is therefore confined to the object rather than emitted from its
surface; the would-be evaporation channel is suppressed by
$\exp(-m_G^{\rm eff}/T)$ relative to ordinary nucleon evaporation and
does not shorten the lifetime of a macroscopic object.  The glueball
portal thus does not open a new drain on quark-nugget dark matter.

\paragraph{Between dense objects: a short-range attraction.}
Exchanged between two such objects, the glueball generates a Yukawa
potential
\beq
V_{G}(r)\simeq\frac{g_{G{\rm obj}}^2}{4\pi}\frac{e^{-m_G r}}{r},
\qquad m_G^{-1}\simeq2\fm\left(\frac{100\MeV}{m_G}\right),
\label{eq:Yukawa}
\eeq
i.e.\ a force of range at most a few fermi --- shorter than the
pion range, not longer.  Two consequences follow, both modest.  First,
the attraction slightly favours the coagulation of small objects into
larger ones, but the potential energy per constituent remains far
below the binding energy that would trigger runaway aggregation; the
objects stay individual.  Second, the contribution to the
self-interaction cross section per unit mass is
\beq
\frac{\sigma}{m}\sim\frac{g_{G{\rm obj}}^4}{4\pi m_G^2 m_{\rm obj}},
\eeq
which for $m_G=100\MeV$ and macroscopic $m_{\rm obj}$ is far below
the Bullet-Cluster bound $\sigma/m\lesssim0.1$--$1 cm^2/{\rm g}$.
Being shorter-ranged than pion exchange, the glueball force cannot
compete with it at nuclear distances, and it does not affect
galaxy-scale dynamics at all.  Its relevance is limited to the
internal dynamics of the densest objects, where the gluon condensate
is enhanced.

\paragraph{One observable consequence.}
If a dense quark object is disrupted --- in a merger, or in the
hot environment of the early Universe --- the trapped glueballs are
released and decay on a microscopic timescale into two photons of
energy $m_G/2$ in the rest frame.  For $m_G=100\MeV$ this is a
quasi-monochromatic soft component at $\sim50\MeV$, Doppler-broadened
by the temperature of the object.  This is the only distinctive
signature of the scenario in a astrophysical setting, and it is a
prediction rather than a constraint: the same coupling that hides the
glueball in the laboratory would reveal it in a merger.

\subsection{Summary}

The ultralight glueball is not dark matter: it carries no conserved
charge and decays on a timescale that is cosmologically
instantaneous, however weakly coupled.  What it does is narrower and
better defined.  During the QCD epoch it acts as a short-lived,
sub-leading dissipative channel that slightly broadens the release of
the trace anomaly and perturbs $N_{\rm eff}$ at the $10^{-2}$ level
for $m_G\sim100\MeV$; it does not alter the crossover itself.  Inside
dense quark objects it is confined rather than emitted, and between
them it mediates a weak attraction of range $1/m_G\lesssim2\fm$ that
is too short-ranged to affect structure formation.  Its one
distinctive observable is a soft, quasi-monochromatic photon line from
the decay of glueballs liberated when such an object is disrupted.
None of these statements requires quark nuggets to exist; they
describe what changes if they do.

\section{Conclusions and discussion}
\label{sec:conclusions}
The energy region between zero and the two-pion threshold,
is not  hadronic desert, may contain important
information about the infrared structure of QCD and the
mechanism of color confinement. The possible existence or
absence of an ultralight glueball in this region 
is the key probe to attack such information.
Existence or absence leads to
 different phenomenological pictures of the low-energy
gluonic sector. 

If such a state exists, it would provide an
independent sub-threshold color-singlet excitation through
which gluonic dynamics could be investigated in full QCD.
We have examined whether an ultralight, flavour- and colour-singlet
glueball,  whose properties were established in the
preceding sections, is compatible with existing data, and whether
any observation could establish or exclude it.  The conclusion is that
current experiments have not excluded it, and that this is a statement
about the coupling rather than about the mass.  Since  the coupling to
nuclear matter can be arbitrarily small, no single measurement can
return a null result that closes the window.  The three avenues we
considered all fail for the same structural reason  for the state does
not couple strongly enough to leave an unambiguous signature.  
Beam dumps and fixed-target searches are
blind to a hadron that deposits a soft, non-resonant cluster.
Forward-cascade and elastic-scattering analyses cannot set a
quantitative bound because the glueball is not singled out among the
hadrons of the same cluster.  The inclusive photon spectra of heavy
quarkonia do reach the kinematic corner in which the glueball would
have to appear, but the published analyses were not designed for it. E.g., 
their $\pi^0$ subtraction is model-dependent, no per-bin covariance was
released, and the last few  MeV below the endpoint are precisely where
the extrapolation is least trustworthy.

The practical consequence is worth stating plainly.  If forthcoming
experiments continue to drive the coupling downward without finding a
signal, one may tentatively conclude that the state is excluded,  but
only in the sense that a coupling which produces no effect at any
sensitivity is indistinguishable from no state at all.  The exclusion
is asymptotic, not finite.

On the positive side we identify two observational avenues that do not
suffer from this ambiguity, because they look for a kinematic
fingerprint rather than a rate.

First, extensive air-shower arrays at ground level.  A glueball
produced in a cosmic-ray interaction and reaching the detector would
manifest itself as an excess of pion-like signals.  It deposits energy
in hadronic channels without the accompanying electromagnetic cascade
that identifies a $\pi^0$.  The observable is therefore a distortion of
the apparent $\pi^0$/charged-pion ratio in the shower.
An excess of this kind would be an indirect indication of glueball
production in cosmic-ray interactions; it would not by itself
distinguish the glueball from other neutral, weakly interacting
hadrons, and it would have to be correlated with the nuclear
saturation constraints of Sec.~\ref{sec:nuclear}.

Second, and more cleanly, the upper endpoint of the photon energy
spectrum in radiative $J/\psi$ and $\Upsilon$ decays.  There the
signal is not an excess in the usual sense but a modification of the
shape: the spectrum must remain non-zero, and smoother, all the way to
the endpoint, especially when  the two gluons can materialise as a state
lighter than the pion.  A coincidence measurement,  a photon within a
 the most hard endpoint paired with a signal in a backward or rear
 calorimeter,  would provide a characteristic fingerprint that no
ordinary background reproduces.  This is in fact unfortunately 
 the very signal that an ``invisible''
analysis vetoes.  As quantified above, the QED
three-photon contribution to this window is suppressed to
$\mathcal{O}(10^{-6})$ or below, and the $\eta$ and $\eta'$ two-body
channels are excluded kinematically; the only conventional backgrounds
able to populate the window are leakage from the $\pi^0$ line.  
Since in this case is that the energetic $\pi^0$ decaying  to photon,
so the crucial difficulty to resolve could only be the precision of the 
photon energy measurement.

\subsection{Discussions: Some topics beyond conclusion}

\paragraph{A heavy-quark dumpling.}

A light color-singlet gluonic cluster may have several possible
roles in hadronic dynamics. If lighter gluonic states with
compatible quantum numbers exist, the cluster may convert
into such states, provided that the corresponding decay
channels are kinematically allowed. It may also undergo
inelastic interactions, transformation, or absorption when
interacting with ordinary hadrons. These possibilities
should be distinguished from spontaneous decay in vacuum.

More intriguingly, the gluonic degrees of freedom associated
with such a cluster may participate in the formation of
compact heavy-quark configurations. Rather than acting merely
as a spectator surrounding a heavy-quark core, they may
influence its color dynamics and provide an additional
mechanism for binding. For example, a fully charmed state
such as $X(6900)$ might involve a $c\bar c c\bar c$ core
coupled to a light color-singlet gluonic excitation. In a
possible decay scenario, the heavy quarks rearrange into two
compact color-singlet $c\bar c$ configurations, which
subsequently emerge as two $J/\psi$ mesons. Their small
color-singlet size may suppress their interaction with the
remaining soft gluonic environment, in a manner reminiscent
of color transparency. Meanwhile, the residual gluonic
degrees of freedom could form a light color-singlet gluonic
state, leading to a channel such as
$X(6900)\to J/\psi J/\psi G$. If this residual state is
sufficiently light and carries sufficiently little energy,
it might escape detection, so that the observed final state
could appear experimentally as a di-$J/\psi$ event.

This construction is a speculative toy picture rather than
an established interpretation of $X(6900)$. Its viability
requires a quantitative understanding of the color dynamics,
the binding mechanism, the relevant decay amplitudes, and
the kinematics and experimental signatures of the residual
gluonic state.

\paragraph{Axion--pseudoscalar glueball mixing.}
The QCD axion couples to the gluonic topological density
through the operator $a\,G\widetilde G/f_a$
\cite{DiLuzio:2020wdo}. A pseudoscalar glueball
$G_{ps}$, with $J^{PC}=0^{-+}$, is associated with the
same gluonic operator. This common QCD origin and the
identical quantum numbers make axion--glueball mixing a
natural possibility to investigate, particularly if their
masses are close. The actual mixing strength is
nonperturbative and model dependent, and would have to be
determined from the relevant gluonic matrix elements.
Such mixing could modify the effective couplings and
decay properties of the physical eigenstates, providing
another possible phenomenological connection between
ultralight gluonic states and axion physics.

\paragraph{The $X(17)$ anomaly and the LZ event.}
The reported $X(17)$ anomaly in nuclear internal pair
conversion, with an invariant mass around $17,\mathrm{MeV}$,
falls within the ultralight mass window considered here
\cite{Krasznahorkay:2015iga,Krasznahorkay:2018gsj,
Krasznahorkay:2021x17,Krasznahorkay:2022x17}.
While a pure scalar glueball cannot directly account for
the observed $e^+e^-$ signal, a predominantly gluonic state
with a small quark or axial/vector admixture may provide
a possible interpretation. Independently, the LUX-ZEPLIN
(LZ) Collaboration has reported a single nuclear-recoil
event at $248\pm 23\,(\mathrm{stat})\pm 23\,(\mathrm{sys})$
keV, in a region with a low expected background
\cite{LZ:2026}. Its origin remains uncertain, and a
connection to an ultralight gluonic state would require
a dedicated analysis of the relevant nucleon couplings.
Neither observation constitutes evidence for an ultralight
glueball at present. Nevertheless, both motivate further
experimental tests of this possibility. If the anomalies
are independently confirmed but cannot be accommodated
within a consistent ultralight-glueball interpretation,
their explanation would point to other forms of new physics.

\appendix

\section{Conformal gluonic continuum: an alternative if the
ultralight glueball is absent}
\label{app:excluded}

The energy region between zero and the two-pion threshold,
although seemingly a hadronic desert, may contain important
information about the infrared structure of QCD and the
mechanism of color confinement. The possible existence or
absence of an ultralight glueball in this region may lead to
very different phenomenological pictures of the low-energy
gluonic sector. If such a state exists, it would provide an
independent sub-threshold color-singlet excitation through
which gluonic dynamics could be investigated in full QCD.
If no such state is found, however, the gluonic degrees of
freedom need not disappear. They may instead manifest
themselves through a continuous infrared structure coupled
to quark degrees of freedom.

This appendix explores the latter possibility as a
phenomenological alternative. It is not intended as a
prediction of standard QCD, nor does the absence of an
ultralight glueball pole by itself imply infrared conformal
invariance. Rather, we investigate whether an approximately
scale-invariant gluonic continuum could provide an effective
description of the low-energy gluonic sector, and consider
some of its possible consequences for hadronic interactions.

\subsection{Motivation and historical precedent}

The idea that a field-theoretic excitation may be physically
relevant without appearing as an ordinary asymptotic particle
has a historical precedent. McCoy and Wu (CERN TH.2487;
Phys.\ Lett.\ B {\bf 72}, 219 (1977)) discussed the notion of
``indeterminate-mass particles'' in connection with the
unobservability of free quarks. In their proposal, the
excitation spectrum need not consist of isolated mass poles,
but could instead exhibit a continuous distribution of
states. Although this particular mechanism for quark
confinement is not the modern description of QCD
confinement, the broader spectral idea remains useful:
field-theoretic degrees of freedom need not always manifest
themselves as isolated particle poles.

Applied to the gluonic sector, the absence of an isolated
ultralight glueball pole below the two-pion threshold would
not imply that gluonic degrees of freedom are absent from
the low-energy theory. One possibility is that their
contribution appears instead through a continuous spectral
distribution, potentially coupled to quark degrees of
freedom. Such a possibility motivates exploring an
unparticle-like description of the infrared gluonic sector.

It is important to distinguish this hypothesis from the
mere absence of a sub-threshold pole. A continuous spectrum
may have a nonzero threshold and therefore coexist with a
mass gap. The scenario considered here is the more specific
possibility that the relevant gluonic spectral distribution
extends down to zero invariant mass and exhibits approximate
scale invariance over some infrared range.

\subsection{An unparticle-inspired infrared description}

Georgi (Phys.\ Rev.\ Lett.\ {\bf 98}, 221601 (2007);
Phys.\ Lett.\ B {\bf 650}, 443 (2007)) introduced an
effective-field-theory framework for sectors that exhibit
approximate scale invariance. Although originally developed
for hypothetical hidden sectors beyond the Standard Model,
the mathematical framework can also be used phenomenologically
to describe a hypothetical approximately scale-invariant
infrared gluonic sector.

The starting point is the possibility that the effective
strong coupling approaches a finite infrared value,
\begin{equation}
\alpha_{\rm eff}(Q^2\to 0)\longrightarrow \alpha_0.
\end{equation}
A finite infrared coupling alone does not establish scale
invariance. For the present scenario, we further assume that
the gluonic sector approaches an approximate infrared fixed
point and that no relevant infrared scale generates a
spectral gap in the channel under consideration.

A suitable gauge-invariant gluonic operator may be chosen,
for example,
\begin{equation}
\mathcal{O}_G^{\rm UV}
=\alpha_s G^a_{\mu\nu}G^{a\mu\nu},
\end{equation}
or an operator with other appropriate quantum numbers.
Its infrared counterpart, denoted by $\mathcal{O}_G$, is
assumed to have a scaling dimension $d_G$ at the hypothetical
infrared fixed point. This infrared scaling dimension need
not coincide with the canonical ultraviolet dimension.

The effective coupling to nucleons can then be parametrized
as
\begin{equation}
\mathcal{L}_{\rm int}^{\rm gl}
=
\frac{c}{\Lambda_{\rm gl}^{d_N+d_G-4}}\,
O_N(x)\,\mathcal{O}_G(x),
\label{eq:app_unparticle_coupling}
\end{equation}
where $O_N$ denotes an appropriate nucleon operator,
such as $\bar N N$, and $\Lambda_{\rm gl}$ is an effective
matching scale. The coefficient $c$ parametrizes the
strength of the coupling and is not assumed to be known
a priori.

For a scalar operator in a scale-invariant sector, the
corresponding spectral density has the characteristic form
\begin{equation}
\rho_G(p^2)
=
A_{d_G}\,
\theta(p^0)\theta(p^2)\,
(p^2)^{d_G-2},
\label{eq:app_spectral}
\end{equation}
where $A_{d_G}$ is a normalization factor. In the limit
$d_G\to 1$, the spectral distribution approaches the
massless-particle pole in the distributional sense,
whereas $d_G>1$ describes a continuous spectrum extending
to zero invariant mass. This is the sense in which the
infrared gluonic excitation may exhibit unparticle-like
behavior. It should not be interpreted as a collection of
freely propagating colored gluons.

\subsection{Gluonic dynamics and quark confinement}

The possible coexistence of an infrared gluonic continuum
and quark confinement requires a distinction between the
dynamics of the gauge field and its coupling to quark
degrees of freedom. In QCD, the Yang--Mills equation in the
presence of quarks is
\begin{equation}
(D_\mu G^{\mu\nu})^a
=
g\,\bar q\gamma^\nu T^a q.
\end{equation}
The left-hand side contains the nonlinear dynamics and
self-interactions of the non-Abelian gauge field, while
the right-hand side represents the quark color current.
These are coupled aspects of the same gauge theory, rather
than independent sectors.

Confinement is a nonperturbative property of this coupled
system. Its descriptions include the area law of Wilson
loops and, in particular effective pictures, the formation
of color flux tubes, center vortices, or magnetic-monopole
configurations. These descriptions need not require the
existence of an independently observable ultralight
glueball pole.

The phenomenological possibility considered here is that
the low-energy gluonic degrees of freedom may remain
dynamically relevant through their coupling to quark
degrees of freedom, even if they do not form an isolated
sub-threshold color-singlet particle. In this sense, the
absence of an ultralight glueball would motivate a
description in terms of a continuous, quark-coupled
infrared gluonic structure.

This possibility does not establish that confinement and
an exactly conformal gluonic sector coexist in QCD.
In particular, a nonzero string tension introduces a
physical scale, and its compatibility with approximate
infrared scale invariance would have to be addressed in a
more complete model. Here, approximate scale invariance
is treated as a phenomenological hypothesis over a
restricted infrared range.

\subsection{Possible consequences for nuclear interactions}

If the infrared gluonic continuum couples to nucleons
through Eq.~\eqref{eq:app_unparticle_coupling}, it may
generate a contribution to the nucleon--nucleon interaction
with a non-Yukawa spatial dependence. For scalar unparticle
exchange, a characteristic scaling form is
\begin{equation}
V_{\rm rem}(r)\sim
\frac{C_{\rm rem}}{\Lambda_{\rm gl}^{2d_G-2}}\,
\frac{1}{r^{2d_G-1}},
\label{eq:app_powerlaw_potential}
\end{equation}
where $C_{\rm rem}$ contains the effective coupling
coefficients and normalization factors. This expression
is intended as a leading scaling estimate in the
appropriate distance regime; contact terms, matching
effects, and the finite range of validity of the
approximate scale-invariant description may modify the
potential.

For example, $d_G=2$ gives a characteristic $1/r^3$
dependence, while the limit $d_G\to1$ approaches the
$1/r$ behavior associated with massless scalar exchange.
A nonzero spectral threshold or an infrared breaking of
scale invariance would modify the long-distance behavior.

The possible contribution of such a continuum would have
to be considered together with the established pion-mediated
nuclear interaction and other short- and intermediate-range
contributions. Consequently, nuclear-force data could
constrain the effective coupling and scaling dimension
of the proposed infrared continuum, provided the assumed
matching to nucleon operators is specified.

The absence of a discrete glueball contribution and the
presence of a continuous gluonic contribution are therefore
phenomenologically distinct possibilities. A search for
sub-threshold glueball states and precision studies of
low-energy nuclear interactions probe complementary aspects
of the infrared gluonic sector.

\subsection{The low-energy gluonic scale anchor}

A sub-threshold glueball, if discovered, would provide a
particularly valuable low-energy spectral anchor for
investigating gluonic dynamics in full QCD. Although
pure-gauge lattice calculations provide an important
theoretical reference for the gluonic spectrum, their
results cannot simply be identified with physical
color-singlet states in full QCD, where mixing with
quark-containing states may occur. The discovery of an
independent ultralight gluonic state would therefore open
a new possibility: the gluonic sector could be studied
directly through a low-mass physical excitation, rather
than only through its indirect effects in mixed hadronic
states and correlation functions.

If no such state is found, this particular spectral
anchor would be absent. Nevertheless, the gluonic
degrees of freedom would still enter the dynamics of
full QCD. The alternative explored in this appendix is
that they may manifest themselves through an infrared
continuum coupled to quark degrees of freedom, possibly
admitting an approximate scale-invariant description.

Thus, the two possible outcomes of a sub-threshold
glueball search have different implications for how
the low-energy gluonic sector should be investigated.
The presence of a discrete state would provide a direct
spectroscopic handle, whereas its absence would motivate
further searches for non-particle-like gluonic structures
and their effects on hadronic observables.

\subsection{Summary}

The absence of an ultralight glueball pole below the
two-pion threshold would not imply that gluonic degrees
of freedom disappear from low-energy QCD. One possible
phenomenological alternative is that they contribute
through a continuous infrared structure coupled to quark
degrees of freedom. If this structure exhibits approximate
scale invariance, an unparticle-inspired description may
provide a useful effective framework.

The main assumptions and implications of this scenario
are:
\begin{enumerate}
\item No isolated ultralight glueball pole is present in
the sub-threshold region under consideration.
\item The relevant gauge-invariant gluonic correlation
functions may nevertheless contain a continuous spectral
contribution extending to very low invariant mass.
\item Over a suitable infrared range, this continuum is
assumed to exhibit approximate scale invariance and may
be parametrized by an effective scaling dimension $d_G$.
\item The gluonic degrees of freedom remain dynamically
coupled to quark degrees of freedom, while confinement
continues to arise from the nonperturbative dynamics of
the full QCD gauge--quark system.
\item If the continuum couples to nucleons, it may produce
additional non-Yukawa contributions to nuclear interactions,
which can in principle be constrained phenomenologically.
\end{enumerate}

This scenario is not a consequence of glueball exclusion,
nor is it presented as an established property of QCD.
It is an alternative infrared possibility whose purpose
is to illustrate that the absence of a discrete
sub-threshold gluonic state would raise a different,
but equally important, question about the physical
manifestation of gluonic degrees of freedom.

The region between zero and the two-pion threshold should
therefore not be regarded as a physically empty hadronic
desert. Whether it contains an independent ultralight
glueball or instead a non-particle-like gluonic continuum,
its structure may provide important clues to the
nonperturbative infrared dynamics of QCD and the mechanism
of color confinement.
```
The region below the two-pion threshold, extending down to zero energy, is far from being a hadronic desert. Whether
the low-energy gluonic sector manifests itself through a
discrete sub-threshold state or a non-particle-like continuum,  it provides an important window into the low-energy gluonic structure of QCD and the mechanism of color confinement.